\documentclass[a4paper,fleqn,usenatbib]{mnras}

\usepackage[T1]{fontenc}
\usepackage{ae,aecompl}

\usepackage{graphicx}	
\usepackage{amsmath}	
\usepackage{amssymb}	

\usepackage[usenames,dvipsnames]{xcolor}
\usepackage{hyperref}
\hypersetup{
    colorlinks = true,
    citecolor = {MidnightBlue},
    linkcolor = {BrickRed},
    urlcolor = {BrickRed}
}

\newcommand\ee{\end{equation}}
\newcommand\be{\begin{equation}}
\newcommand\eea{\end{eqnarray}}
\newcommand\bea{\begin{eqnarray}}

\newcommand\lsim{\mathrel{\rlap{\lower4pt\hbox{\hskip1pt$\sim$}}
        \raise1pt\hbox{$<$}}}
\newcommand\gsim{\mathrel{\rlap{\lower4pt\hbox{\hskip1pt$\sim$}}
        \raise1pt\hbox{$>$}}}

\newcommand{\bn}{\mathbf{n}}

\newcommand{\bV}{\mathbf{V}} 
\newcommand{\bx}{\mathbf{x}}
\newcommand{\by}{\mathbf{y}}

\newcommand{\bk}{\mathbf{k}}

\newcommand{\HH}{{\cal H}}

\newcommand{\dd}{\partial}

\newcommand{\g}{{\rm g}}
\newcommand{\vv}{{\rm v}}

\newcommand{\dmin}{d_{\rm min}}
\newcommand{\dmax}{d_{\rm max}}

\newcommand{\norm}{a_{\rm N}}

\title[Doppler magnification dipole]{Dipolar modulation in the size of galaxies:\\ The effect of Doppler magnification}

\author[C. Bonvin et al.]{Camille Bonvin$^1$\thanks{camille.bonvin@unige.ch},
Sambatra Andrianomena$^{2,3,4}$\thanks{andrianomena@gmail.com},
David Bacon$^5$\thanks{david.bacon@port.ac.uk},
\newauthor
Chris Clarkson$^{2,6}$\thanks{chris.clarkson@qmul.ac.uk},
Roy Maartens$^{3,5}$\thanks{roy.maartens@gmail.com},
Teboho Moloi$^2$\thanks{tebzanaaka@gmail.com},
Philip Bull$^{7,8}$\thanks{philbull@gmail.com}
\\
${}^{1}$D\'epartement de Physique Th\'eorique and Center for Astroparticle Physics (CAP), \\
$\phantom{x}$University of Geneva, 24 quai Ernest Ansermet, CH-1211 Geneva, Switzerland.\\
${}^2$Department of Mathematics \& Applied Mathematics, University of Cape Town, Cape Town 7701, South Africa.\\
${}^{3}$Department of Physics \& Astronomy, University of the Western Cape, Cape Town 7535, South Africa.\\
${}^4$SKA South Africa, 3rd Floor, The Park, Park Road, Pinelands, 7405, South Africa.\\
${}^5$Institute of Cosmology and Gravitation, University of Portsmouth, Portsmouth, PO1 3FX, UK.\\
${}^{6}$School of Physics \& Astronomy, Queen Mary University of London, Mile End Road, London E1 4NS, UK.\\
${}^7$California Institute of Technology, Pasadena, CA 91125, USA.\\
${}^8$Jet Propulsion Laboratory, California Institute of Technology, 4800 Oak Grove Drive, Pasadena, California, USA.}

\date{Accepted XXX. Received YYY; in original form ZZZ}

\pubyear{2016}

\begin{document}
\label{firstpage}
\pagerange{\pageref{firstpage}--\pageref{lastpage}}
\maketitle

\begin{abstract} 

Objects falling into an overdensity appear larger on its near side and smaller on its far side than other objects at the same redshift. This produces a dipolar pattern of magnification, primarily as a  consequence of the Doppler effect. At low redshift this Doppler magnification completely dominates the usual integrated gravitational lensing contribution to the lensing magnification. We show that one can optimally observe this pattern by extracting the dipole in the cross-correlation of number counts and galaxy sizes. This dipole allows us to almost completely remove the contribution from gravitational lensing up to redshift $\lsim 0.5$, and even at high redshift $z\simeq 1$ the dipole picks up the Doppler magnification predominantly. Doppler magnification should be easily detectable in current and upcoming optical and radio surveys; by forecasting for telescopes such as the SKA, we show that this technique is competitive with using peculiar velocities via redshift-space distortions to constrain dark energy. It produces similar yet complementary constraints on the cosmological model to those found using measurements of the cosmic shear. 

\end{abstract}


\section{Introduction}
\label{sec:intro}

Gravitational lensing is a powerful cosmological probe that is sensitive to the distribution of matter between the source and the observer. It provides a measurement of the gravitational potentials integrated along the photon trajectory and is therefore sensitive to the growth rate of structure. Gravitational lensing can be measured through two distinct observables: the shear and the convergence. The shear $\gamma$ encodes the effect of lensing on the observed shape of galaxies. An estimator for the shear can be constructed from the ellipticity of galaxies. The convergence $\kappa$ accounts for the effect of lensing on the observed size of galaxies. The convergence is in principle more difficult to measure than the shear, since the mean intrinsic size of galaxies at a given redshift is unknown, whereas the mean intrinsic ellipticity is expected to vanish. Recently an estimator for the convergence has been proposed, combining measurements of the size and magnitude of galaxies~\citep{2012ApJ...744L..22S, Casaponsa:2012tq, Heavens:2013gol, Alsing:2014fya}. The signal-to-noise of the convergence using this estimator has been shown to be about half of that of the shear. Since the convergence is affected by different systematics than the shear, this estimator provides a valuable complementary tool to measure gravitational lensing.  

However, it has also been shown that, unlike the shear, the convergence is not only affected by gravitational lensing but also by various other effects: Doppler effects, Sachs-Wolfe effects, Shapiro time-delay and the integrated Sachs-Wolfe effect~\citep{Bonvin:2008ni, Bolejko:2012uj, Bacon:2014uja}. The physical origin of these effects is easy to understand: they modify the apparent distance between the observer and the galaxies at a given redshift and consequently they change their observed size. The most important of these contributions at sub-horizon scales is the Doppler correction due to the peculiar velocity of galaxies. This effect has been called {\it Doppler magnification} (or Doppler lensing). 

The fact that galaxy peculiar velocities affect the observed distance to objects is well known, and has been extensively used in the context of nearby objects to measure the velocity~\citep[see e.g.][and the discussion in Section~\ref{sec:distance}]{1977A&A....54..661T, Dressler:1987ny, Djorgovski:1987vx, Tonry:1999vt, Turnbull:2011ty, Tully:2013wqa, Springob:2014qja}. The low redshift expression relating the distance to the velocity has subsequently been extended to higher redshift and general expressions for the fluctuations in the luminosity distance (which can easily be related to the convergence) have been derived~\citep{1987MNRAS.228..653S, 1989PhRvD..40.2502F, 1990PhLA..147...97K, Pyne:2003bn, Bonvin:2005ps, Hui:2005nm, Kaiser:2014jca}. However this Doppler magnification is usually not accounted for in weak lensing analyses. The reason is twofold: first the Doppler magnification affects only the size of galaxies at linear order, but it leaves their shape unchanged. As a consequence the cosmic shear $\gamma$, which is the primary target of lensing surveys, is not affected by Doppler lensing at linear order~\footnote{Note that at second-order in perturbation theory, this effect contributes to the shear in a non-negligible way~\citep{Bernardeau:2009bm, Bernardeau:2011tc}.}. The second reason why Doppler magnification is usually neglected in lensing analyses is because it  becomes subdominant with respect to gravitational lensing as the redshift increases. This is because gravitational lensing accumulates along the line-of-sight whereas the Doppler magnification is a local effect which decreases with redshift. Measurements of $\langle \kappa(\bn) \kappa(\bn')\rangle$ at redshift larger than $\sim 0.5$ are therefore relatively insensitive to Doppler magnification~\citep{Bonvin:2008ni}.

\citet{Bacon:2014uja} proposed a new method to detect the Doppler magnification by cross-correlating the convergence $\kappa$, estimated through galaxy sizes and magnitudes, with the galaxy number count contrast $\Delta$. As shown there, the two-point function $\langle \Delta\kappa\rangle$ is anti-symmetric around $\Delta$. Bacon et al. constructed  an estimator, based on the angular power spectrum $C_\ell$, which changes sign when the measured convergence  is in front of or behind the density contrast $\Delta$, to target the Doppler magnification. They showed that this can be used to constrain the cosmological model, and also to reconstruct the peculiar velocity field on cosmological scales.

In this paper we propose an improved estimator in configuration space that allows us to optimally exploit the anti-symmetry of the two-point function $\langle \Delta\kappa\rangle$. We use the formalism developed for redshift-space distortion measurements, i.e. we associate to each pair of pixels $(i, j)$ a separation $d_{ij}$ and an orientation with respect to the line-of-sight $\beta_{ij}$ (see Figure~\ref{fig:coordinate}). In one of those pixels we measure the galaxy number count $\Delta_i$ and in the other we measure the convergence $\kappa_j$. We then expand the mean of the two-point function in Legendre polynomial and show that the Doppler magnification induces a dipole and an octupole. Consequently we propose the following estimators to optimally measure Doppler magnification
\bea
\xi_{\rm dip}(d)&=&\norm \sum_{ij} \Delta_i \kappa_j \cos\beta_{ij}\delta_K(d_{ij}-d)\, ,\\
\xi_{\rm oct}(d)&=&b_{\rm N} \sum_{ij} \Delta_i \kappa_j P_3(\cos\beta_{ij})\delta_K(d_{ij}-d)\, ,
\eea
where $\norm$ and $b_{\rm N}$ are normalisation factors and $P_3$ is the Legendre polynomial of degree 3.
We show that the dipole estimator allows us to almost completely remove the contribution from gravitational lensing up to redshift $\sim 0.5$, and that even at high redshift $z\simeq 1$ the dipole picks up the Doppler magnification predominantly. It therefore provides a new way of measuring peculiar velocities by observing the size of galaxies. We then calculate the signal-to-noise of the dipole and the octupole in a selection of optical and radio surveys. Depending on the error associated with the measurement of the convergence, we find a cumulative signal-to-noise of $12-31$ (the first number is associated with a size error of $\sigma_\kappa=0.8$ and the second is for $\sigma_\kappa=0.3$~\citep{Alsing:2014fya}) combining the main sample of SDSS, the LOWz and the CMASS samples. For the upcoming DESI bright galaxy sample~\citep{Levi:2013gra}, along with imaging, we forecast a signal-to-noise of $14-37$. For SKA Phase 2, combining redshifts $0.1\leq z\leq 0.5$ we find a cumulative signal-to-noise of $35-93$. The octupole is significantly smaller than the dipole and cannot be detected in current optical surveys. For DESI we find however a cumulative signal-to-noise of $1.9-5$ and for the SKA $5.1-14$. This demonstrates the detectability of Doppler magnification in both optical and radio surveys. We then perform a Fisher forecast analysis and show that the Doppler magnification can provide constraints on cosmological parameters which are competitive with standard redshift-space distortion measurements. 

The remainder of the paper is organised as follows: in Section~\ref{sec:correlation} we derive the general form of the cross-correlation between $\Delta$ and $\kappa$. In Section~\ref{sec:estimator}, we construct an estimator to measure the dipole and the octupole generated by the Doppler magnification. We discuss the contamination from gravitational lensing and the importance of wide-angle and evolution effects. In Section~\ref{sec:variance} we calculate the variance of our estimator and compute the signal-to-noise in optical and radio surveys. We present Fisher forecasts in Section~\ref{sec:fisher}, and compare with other velocity estimators in Section~\ref{sec:comp}. Finally, we conclude in Section~\ref{sec:conclusion}.   

\section{Multipole expansion of the cross-correlation $\langle \Delta \kappa\rangle$}
\label{sec:correlation}

We shall consider the cross-correlation between the number count contrast of galaxies $\Delta$ and the convergence $\kappa$
\be
\xi=\langle \Delta(z,\bn)\kappa(z',\bn') \rangle\, ,
\ee
where $z$ denotes the redshift and $\bn$ the direction of observation. The number count contrast of galaxies can be written as~\citep{Yoo:2009au, Yoo:2010ni, Bonvin:2011bg, Challinor:2011bk, Jeong:2011as}
\begin{align}
\label{Delta}
\Delta(z, \bn)=&b\,\delta-\frac{1}{\HH}\partial_r(\bV\cdot\bn) \\
&+(5s-2)\int_0^{r}\!dr'\,\frac{r-r'}{2rr'}\Delta_\Omega(\Phi+\Psi)+\Delta^{\rm rel}(z,\bn)\, ,\nonumber
\end{align}
where $\Phi$ and $\Psi$ are the two metric potentials\footnote{We use here the following convention for the metric $ds^2=a^2\big[-(1+2\Psi)d\eta^2+(1-2\Phi)\delta_{ij}dx^i dx^j \big]$, where $a$ is the scale factor and $\eta$ denotes conformal time.}, $\bV$ is the peculiar velocity,  $\HH$ is the conformal Hubble parameter, $r$ is the conformal distance to the source, $b$ and $s$ denote respectively the bias and the slope of the luminosity function and the operator $\Delta_\Omega$ is the angular part of the Laplacian
\be
\Delta_\Omega=r^2\left(\nabla^2- n^i n^j\partial_i\partial_j-\frac{2}{r} n^i\partial_i\right)\, .
\ee
The first term in \eqref{Delta} represents the contribution from dark matter density fluctuations (assuming a linear galaxy bias), the second term is the well-known redshift-space distortions, the third term denotes the lensing magnification bias and the last term $\Delta^{\rm rel}$ encodes the so-called relativistic distortions.

The general expression for the convergence at linear order is given by~\citep{Bonvin:2008ni, Bolejko:2012uj, Bacon:2014uja} 
\begin{align}
\kappa(z,\bn)=&\frac{1}{2r}\int_{0}^{r} dr'\frac{r-r'}{r'}\Delta_\Omega\big(\Phi+\Psi\big)
+\left(\frac{1}{r\HH} -1\right)\bV\cdot\bn  \nonumber\\
&-\frac{1}{r}\int_0^r dr'\big(\Phi+ \Psi\big)+\left(1-\frac{1}{r\HH} \right)\int_0^r dr'\big(\dot{\Phi}+\dot\Psi\big)\nonumber\\
&+\left(1-\frac{1}{r\HH} \right)\Psi+\Phi \label{kappa}\, .
\end{align}
In addition to the standard gravitational lensing contribution (first term), we see that the convergence contains a Doppler magnification (second term), a Shapiro time-delay and an integrated Sachs-Wolfe (second line) and a Sachs-Wolfe contribution (third line). Note that we neglect the contributions to $\Delta$ and $\kappa$ at the observer position. The terms proportional to the gravitational potentials at the observer, $\Phi_O$ and $\Psi_O$, contribute only to the local monopole around the observer, and so are always subtracted observationally ($\Delta$ and $\kappa$ are defined as the difference with respect to the total mean). In addition, the contributions proportional to the peculiar velocity at the observer, $\bV_O\cdot\bn$, generate a local dipole around the observer, which can easily be fitted for and subtracted from the perturbations, as done in CMB analyses for example.\footnote{Note that even if the local dipole is not subtracted from the perturbations, its contribution to our estimator should be negligible. In the distant-observer approximation, the velocity of the observer affects all galaxies in the same way (since in this case $\bn=\bn'$) and therefore this contribution exactly vanishes when we fit for a dipole and an octupole around $\Delta$; see Eqs.~\eqref{estimatordip} and~\eqref{estimatoroct}. In the full-sky limit, a small contribution may remain, however.} 

The cross-correlation between $\Delta$ and $\kappa$ contains a large number of terms. In this paper we concentrate on the dominant contributions, given by
\be
\xi=\left\langle \left(b\, \delta -\frac{1}{\HH} \dd_r(\bV\cdot\bn)\right) \Big(\kappa_\g+\kappa_\vv\Big)\right\rangle=\xi_\g+\xi_\vv\, ,
\ee
where $\kappa_\g$ and $\kappa_\vv$ denote respectively the gravitational lensing contribution and the Doppler magnification
\begin{align}
\kappa_\g&=\frac{1}{2r}\int_{0}^{r} dr'\frac{r-r'}{r'}\Delta_\Omega(\Phi+\Psi)\, ,\\
\kappa_\vv&=\left(\frac{1}{r\HH} -1\right)\bV\cdot\bn\, , \label{kappav}
\end{align}
and $\xi_\g$ and $\xi_\vv$ denote the individual $\Delta$-$\kappa_\g$ and  $\Delta$-$\kappa_\vv$ cross-corelations.

\subsection{Doppler magnification}

Let us start by calculating the Doppler magnification contribution to $\xi$. Using the Fourier transform convention 
\be
f(\bx, \eta)=\frac{1}{(2\pi)^3}\int d^3\bk\, e^{-i\bk\cdot\bx}f(\bk,\eta)\, ,
\ee
we can express the cross-correlation as
\begin{align}
&\xi_\vv(z,z',\theta)=i\frac{\HH(z')}{\HH_0}f(z')\left(\frac{1}{\HH(z')r'}-1 \right)\int \frac{d^3\bk}{(2\pi)^3}\,e^{i\bk\cdot(\bx-\bx')}\nonumber\\
&\times P(k, z, z')\frac{\HH_0}{k}
\left[b(z)+\frac{f(z)}{3}+\frac{2f(z)}{3}P_2(\hat\bk\cdot\bn)\right] P_1(\hat\bk\cdot\bn')\, , \label{cross}
\end{align}
where $f={d\ln D}/{d \ln a}$ denotes the growth rate ($D$ is the growth function), $P_1(x)=x$ and $P_2(x)=(3x^2-1)/2$ are the Legendre polynomials of order 1 and 2 and $P(k, z, z')$ is the matter power spectrum defined through
\be
\langle\delta(\bk, z)\delta(\bk', z') \rangle=(2\pi)^3 P(k, z, z') \delta_D(\bk+\bk')\, .
\ee 
\begin{figure}
\centering
\includegraphics[width=0.23\textwidth]{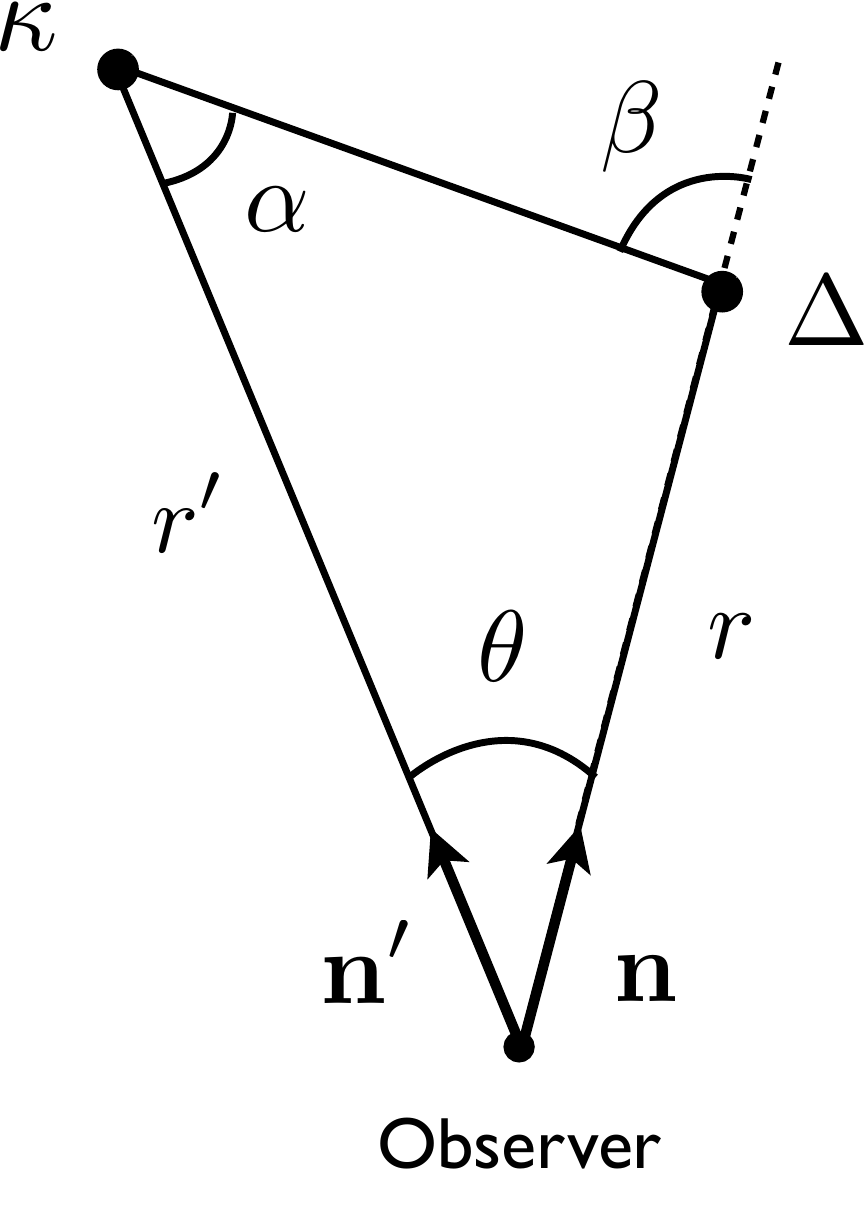}
\caption{\label{fig:coordinate} Coordinate system: the cross-correlation can be expressed in terms of three variables. Two common choices are $(r, r', \theta)$ (or equivalently $(z, z', \theta)$, see footnote 3); and $(r, d, \beta)$ (or equivalently $(z, d, \beta)$).}
\end{figure}
The cross-correlation~\eqref{cross} is a function of  $(z, z', \theta)$, where $\theta$ is the angle between $\bn$ and $\bn'$. We can re-express this cross-correlation in terms of $(z, d, \beta)$, where $d$ is the comoving distance between the galaxies and $\beta$ is the orientation of the pair with respect to the line-of-sight (see Figure~\ref{fig:coordinate}). Following~\citet{Szalay:1997cc, Szapudi:2004gh, Papai:2008bd, Montanari:2012me}, we expand the exponential and Legendre polynomials in terms of spherical harmonics, which allows us to integrate over the direction of $\bk$. The cross-correlation then takes the simple form (see also Appendix B of~\citet{Bonvin:2013ogt} for a similar detailed derivation)
\begin{align}
\xi_\vv(z,z',\beta)=&\frac{\HH(z')}{\HH_0}f(z')\left(1-\frac{1}{\HH(z')r'} \right)\label{xifull}\\
\times& \Bigg\{\left[\left(b(z)+\frac{2f(z)}{5} \right)\nu_1(d)-\frac{f(z)}{10}\nu_3(d) \right]\cos\alpha\nonumber\\
& +\frac{f(z)}{5}\Big[\nu_1(d)-\frac{3}{2}\nu_3(d) \Big]\cos\alpha\cos2\beta\nonumber\\
&+\frac{f(z)}{5}\Big[\nu_1(d)+\nu_3(d) \Big]\sin\alpha\sin2\beta\Bigg\}\, , \nonumber
\end{align}
where
\be
\label{nuell}
\nu_\ell(d)=\frac{1}{2\pi^2}\int dk k^2\frac{\HH_0}{k}P(k, z, z')j_\ell(kd)\,, \hspace{0.1cm} \ell=1,3\, ,
\ee
and $j_\ell$ are the spherical Bessel functions~\footnote{Note that here we use $z$ and $r$ (and similarly $z'$ and $r'$) interchangeably since they are related by their background relation, $1+z(r)=1/a(r)$. The corrections induced by the fluctuations in the redshift have already been consistently included in the expressions for $\Delta$ and $\kappa$, Eqs.~\eqref{Delta} and~\eqref{kappa}.}.
The comoving distance to $\kappa$, $r'$, and the angle $\alpha$ can be explicitly written in terms of $(r, d, \beta)$:
\begin{align}
r'&=\sqrt{r^2+2dr\cos\beta+d^2}\, , \label{rprime}\\
\cos\alpha&=\frac{d+r\cos\beta}{\sqrt{r^2+2dr\cos\beta+d^2}}\, , \label{cosalpha}\\
\sin\alpha&=\frac{r\sin\beta}{\sqrt{r^2+2dr\cos\beta+d^2}}\, . \label{sinalpha}
\end{align}
Eqs.~\eqref{xifull} to \eqref{sinalpha} provide the general linear expression (valid at all scales) for the cross-correlation between the galaxy number counts and the Doppler magnification, as a function of the three variables $(r, d, \beta)$.
These expressions can be further simplified in the distant observer approximation, i.e. in the regime where $d/r\ll 1$. In this limit, we have
\begin{align}
r'&=r+\mathcal{O}\left(\frac{d}{r} \right)\, ,\\
\cos\alpha&=\cos\beta +\mathcal{O}\left(\frac{d}{r} \right)\, ,\\
\sin\alpha&= \sin\beta+\mathcal{O}\left(\frac{d}{r} \right)\, . 
\end{align}
Moreover, all functions of $z'\equiv z(r')$ can be Taylor expanded around $r$. For example, the Hubble parameter $\HH(z')$ becomes at lowest order in $d/r$
\be
\HH(z')=\HH(r')=\HH(r)+\mathcal{O}\left(\frac{d}{r} \right)\, , \label{Hevol}
\ee
and similarly for $f(z')$ and $P(k, z, z')$. With these approximations, Eq.~\eqref{xifull} becomes, at lowest order in $d/r$,
\begin{align}
&\xi_\vv(r,d,\beta)=\frac{\HH(z)}{\HH_0}f(z)\left(1-\frac{1}{\HH(z)r} \right)\label{xiv}\\
&\times\left\{\left(b(z)+\frac{3f(z)}{5} \right)\nu_1(d)P_1(\cos\beta)-\frac{2f(z)}{5}\nu_3(d)P_3(\cos\beta) \right\} \nonumber
\end{align}
In the distant observer approximation, the cross-correlation between the galaxy number counts and the Doppler magnification can therefore be expressed as the sum of a dipole (proportional to $P_1(\cos\beta)$), and an octupole (proportional to $P_3(\cos\beta)$). The cross-correlation is completely anti-symmetric: it changes sign when the convergence is evaluated in front of or behind the overdensity (i.e. when $\beta\to\pi-\beta$). This can be intuitively understood by noting that galaxies tend to move towards overdense regions. On average, galaxies in front of overdensities are therefore moving away from the observer and are apparently magnified by the Doppler magnification term, whereas galaxies behind overdensities are moving towards the observer and are apparently demagnified.

\subsection{Gravitational lensing}

The cross-correlation between the gravitational lensing contribution $\kappa_{\rm g}$ and the galaxy number counts is also expected to have an asymmetric contribution: galaxies behind an overdense region will be magnified by it, whereas galaxies in front of an overdense region will not be affected. This cross-correlation can be calculated using the Limber approximation. It reads
\begin{align}
\xi_\g(r, d, \beta)&=\frac{3 \Omega_m}{2 a\pi} b(z)\frac{r(r'-r)}{2r'}\Theta(r'-r)\label{xilens}\\
&\times \int_0^\infty dk_\perp k_\perp\HH_0P(k_\perp, z,z')J_0(k_\perp |\Delta \bx_\perp|) \, ,\nonumber
\end{align}
where $|\Delta \bx_\perp|=d |\sin\beta|$ 
is the transverse separation between $\bx$ and $\bx'$, $k_\perp$ is the transverse component of the wavenumber and $\Theta(y)$ is the Heaviside function: $\Theta(y)=1$ if $y>0$ and zero elsewhere. We see that in the Limber approximation, the cross-correlation is therefore non-zero only if $r'>r$, i.e. when the convergence is evaluated behind an overdensity. The dependence of Eq.~\eqref{xilens} on the angle $\beta$ is non-trivial, since it is given not only by the pre-factor
\be
\frac{r'-r}{r'}=\frac{d}{r}\cos\beta+\mathcal{O}\left(\frac{d}{r} \right)^2\, ,
\ee 
but also by the argument of the Bessel function $J_0$. Therefore, even in the flat-sky approximation, the cross-correlation between gravitational lensing and the galaxy number count cannot be expressed analytically as a simple multipole expansion. The multipoles can however be calculated numerically, by weighting the cross-correlation by the appropriate Legendre polynomial.

\section{Estimator}
\label{sec:estimator}

\begin{figure*}
\centering
\includegraphics[width=0.48\textwidth]{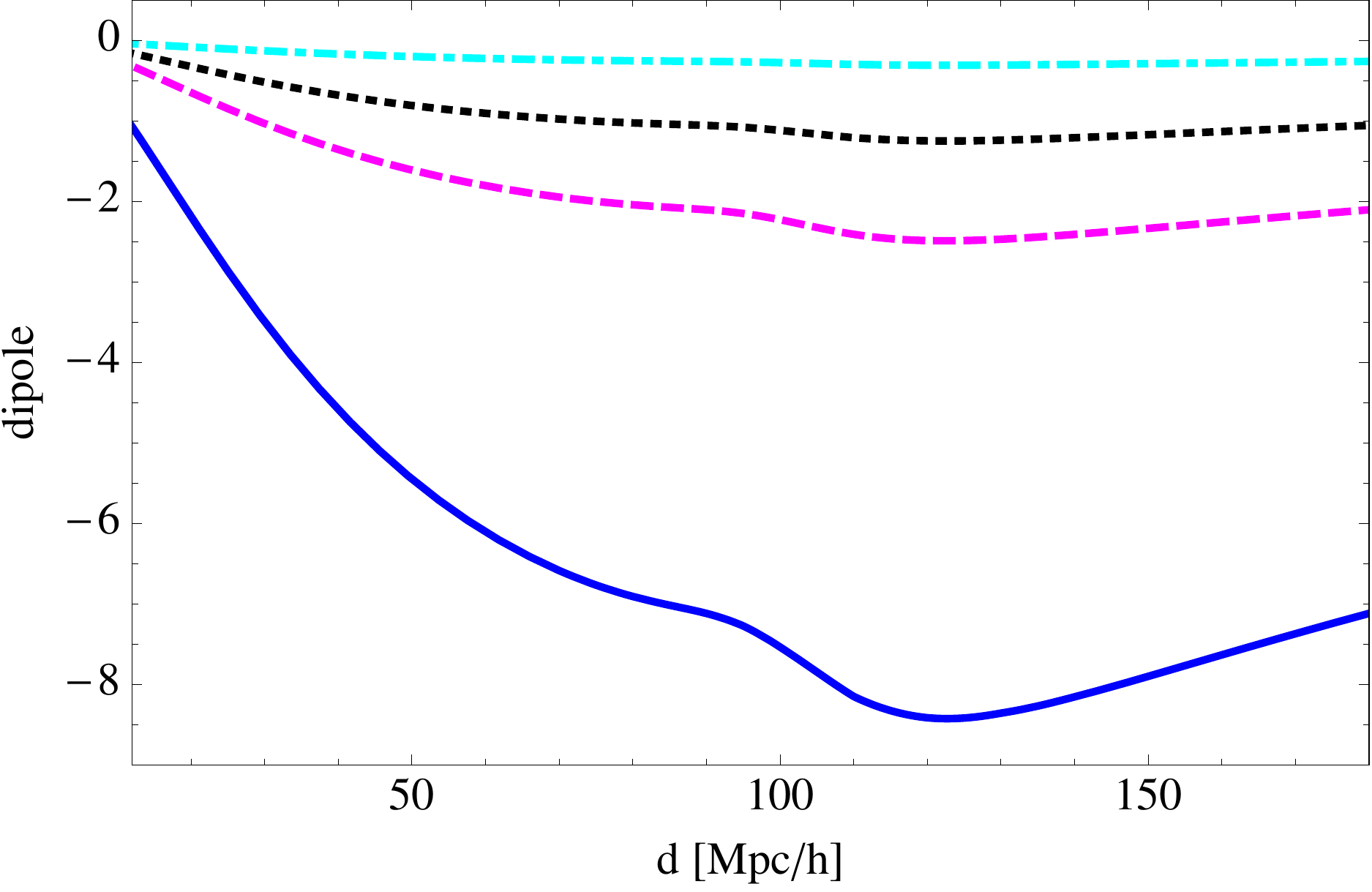}\hspace{0.5cm}\includegraphics[width=0.48\textwidth]{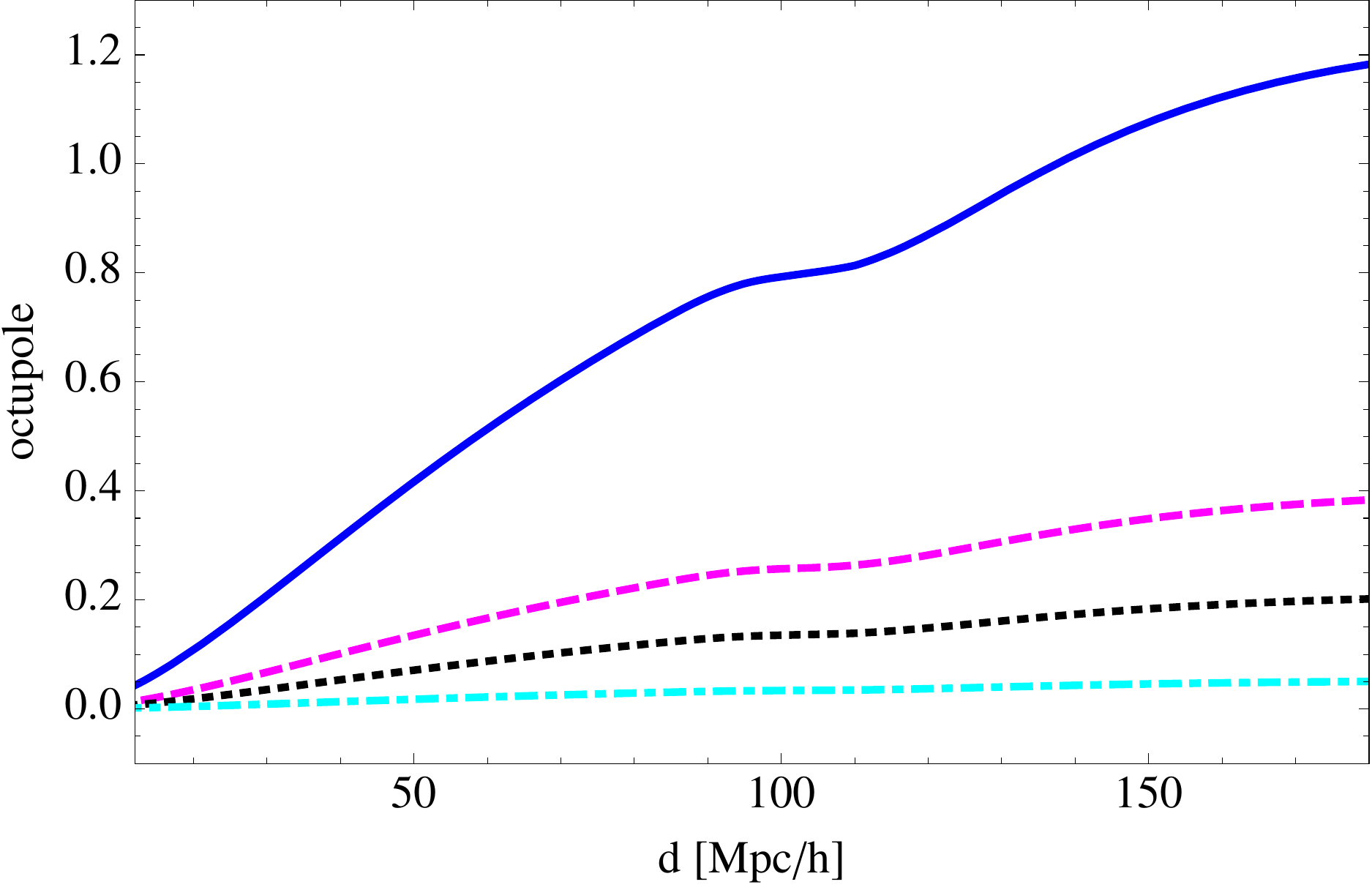}
\caption{\label{fig:dipall} {\it Left panel}: Amplitude of the Doppler magnification dipole~\eqref{meandip}, plotted as a function of separation $d$, at four different redshifts: $z=0.1$ (blue solid), $z=0.3$ (magenta dashed), $z=0.5$ (black dotted) and $z=1$ (cyan dash-dotted).
{\it Right panel}: Amplitude of the Doppler lensing octupole~\eqref{meanoct} at the same four redshifts. The dipole and octupole are multiplied by $d^2$. We show scales between 12\,Mpc/$h$ and 180\,Mpc/$h$, that will be used in the Fisher forecasts.}
\end{figure*}

\begin{figure*}
\centering
\includegraphics[width=0.48\textwidth]{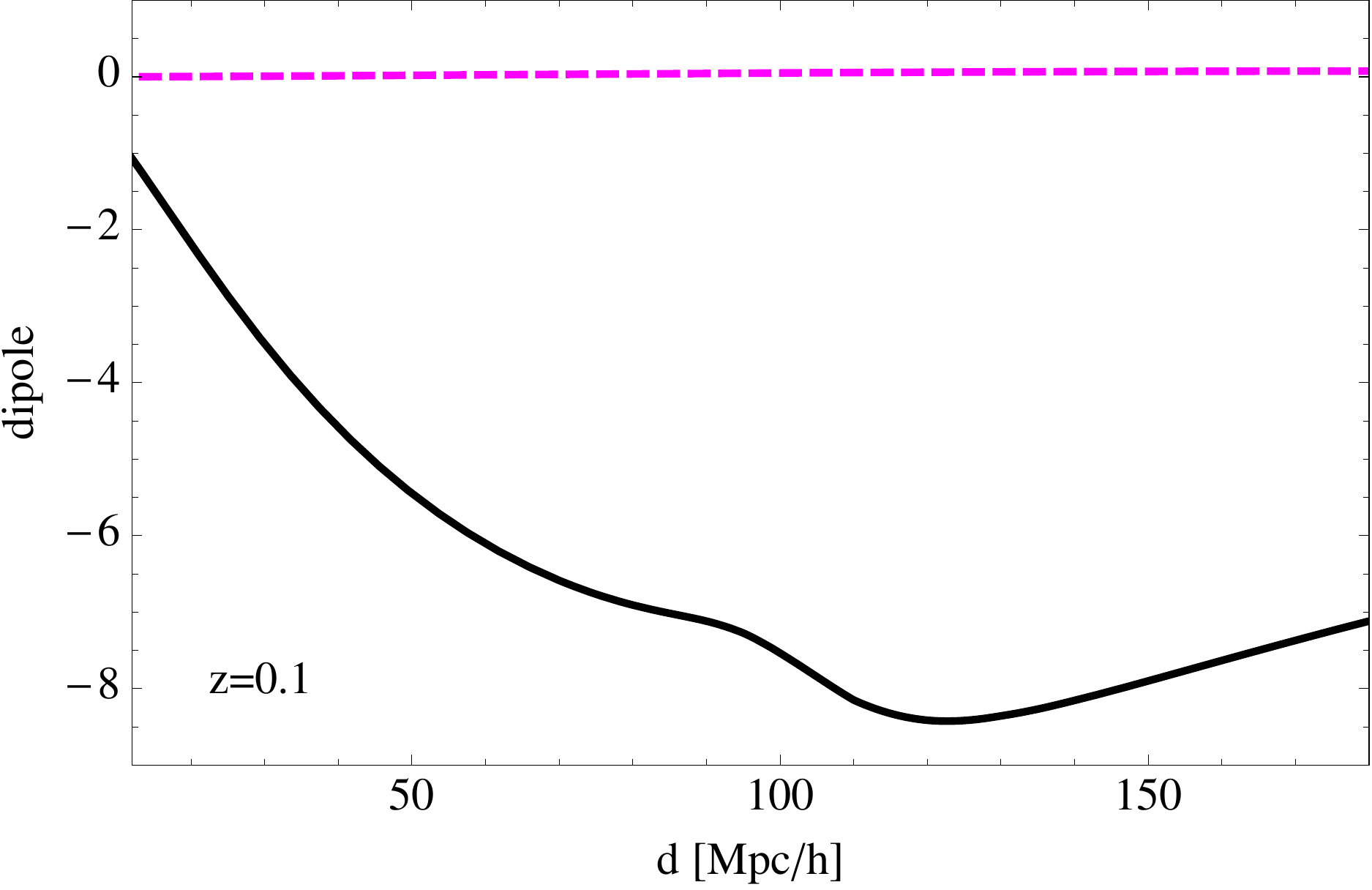}\hspace{0.5cm}\includegraphics[width=0.48\textwidth]{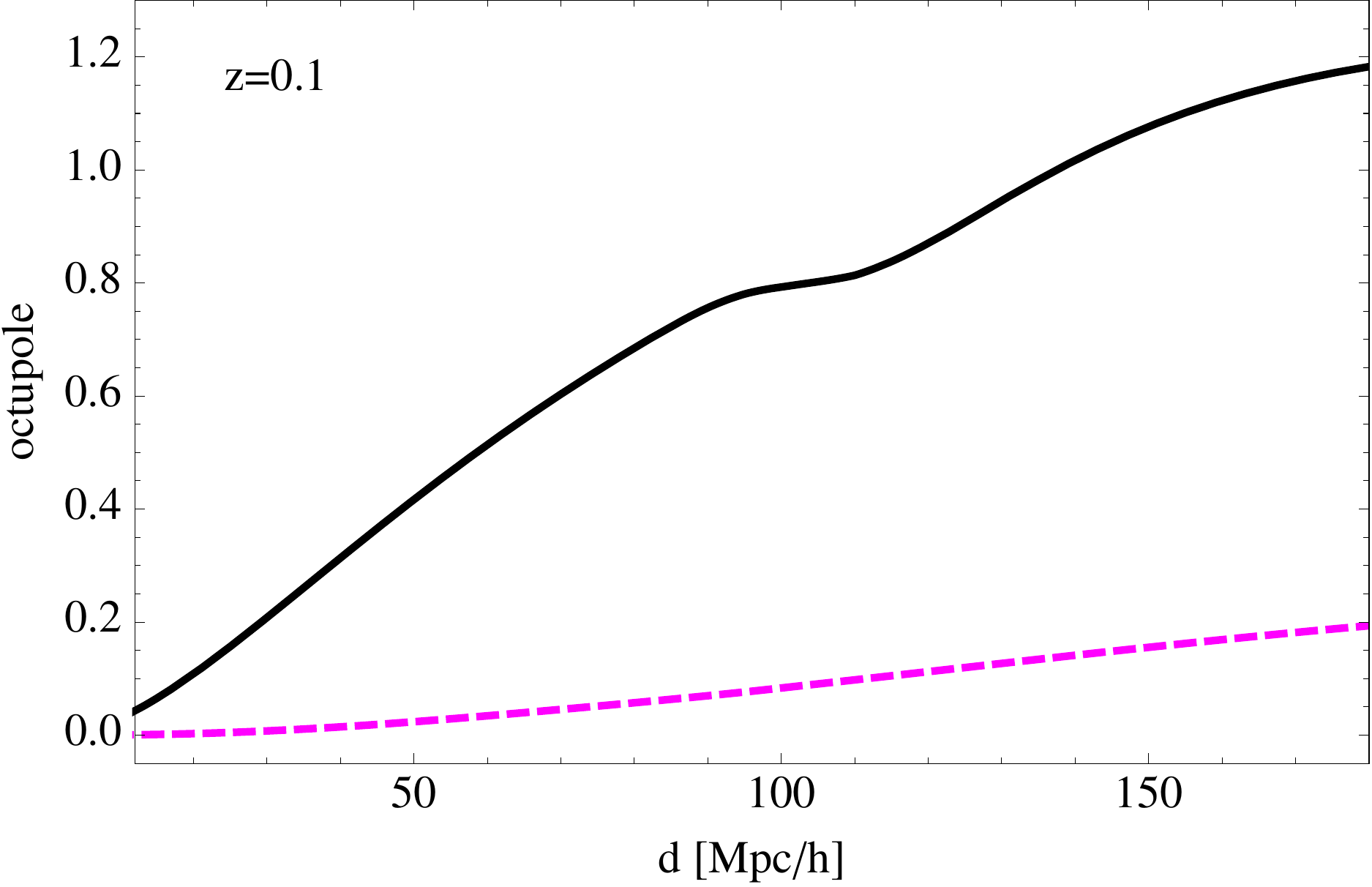}\\
\includegraphics[width=0.48\textwidth]{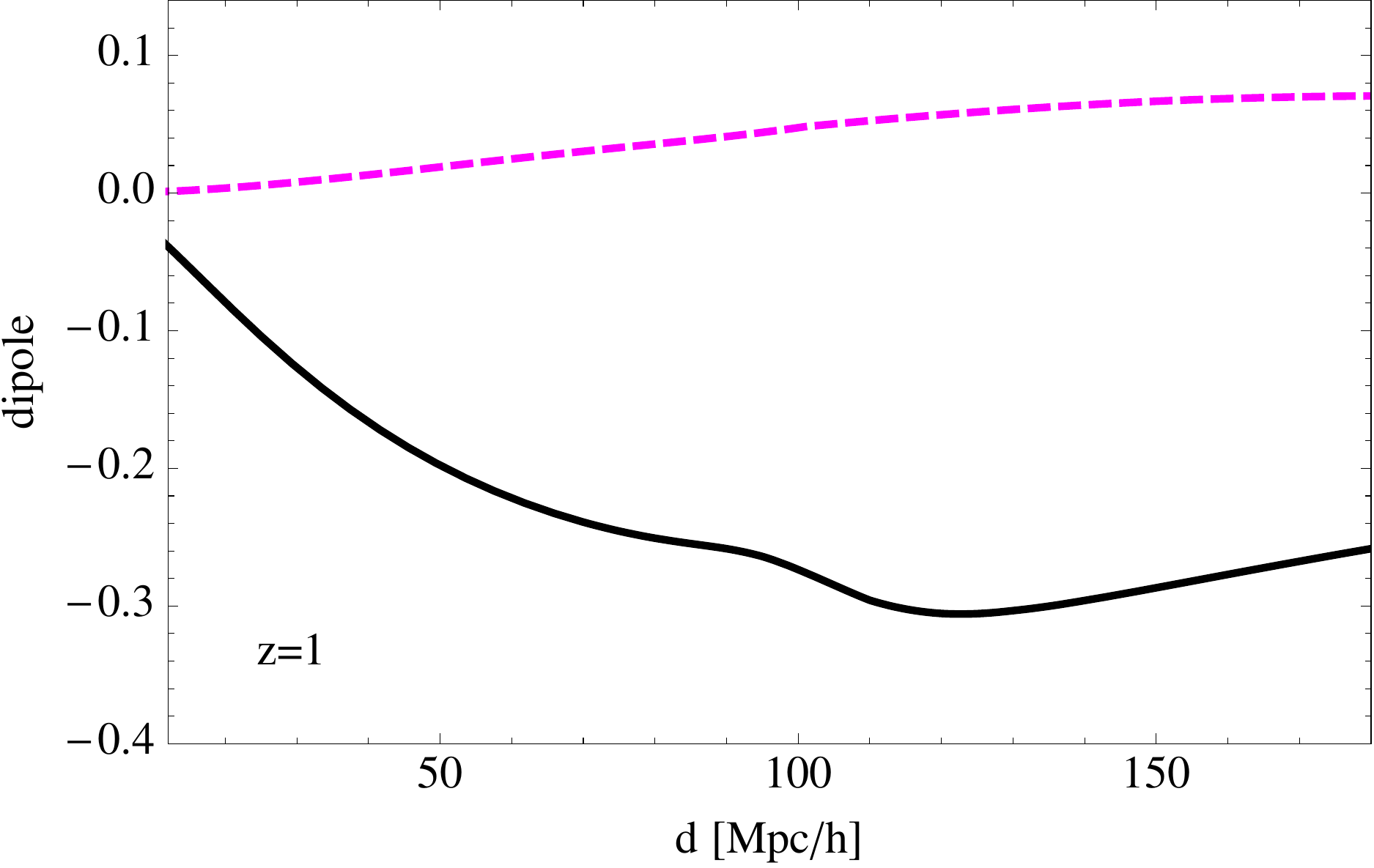}\hspace{0.5cm}\includegraphics[width=0.48\textwidth]{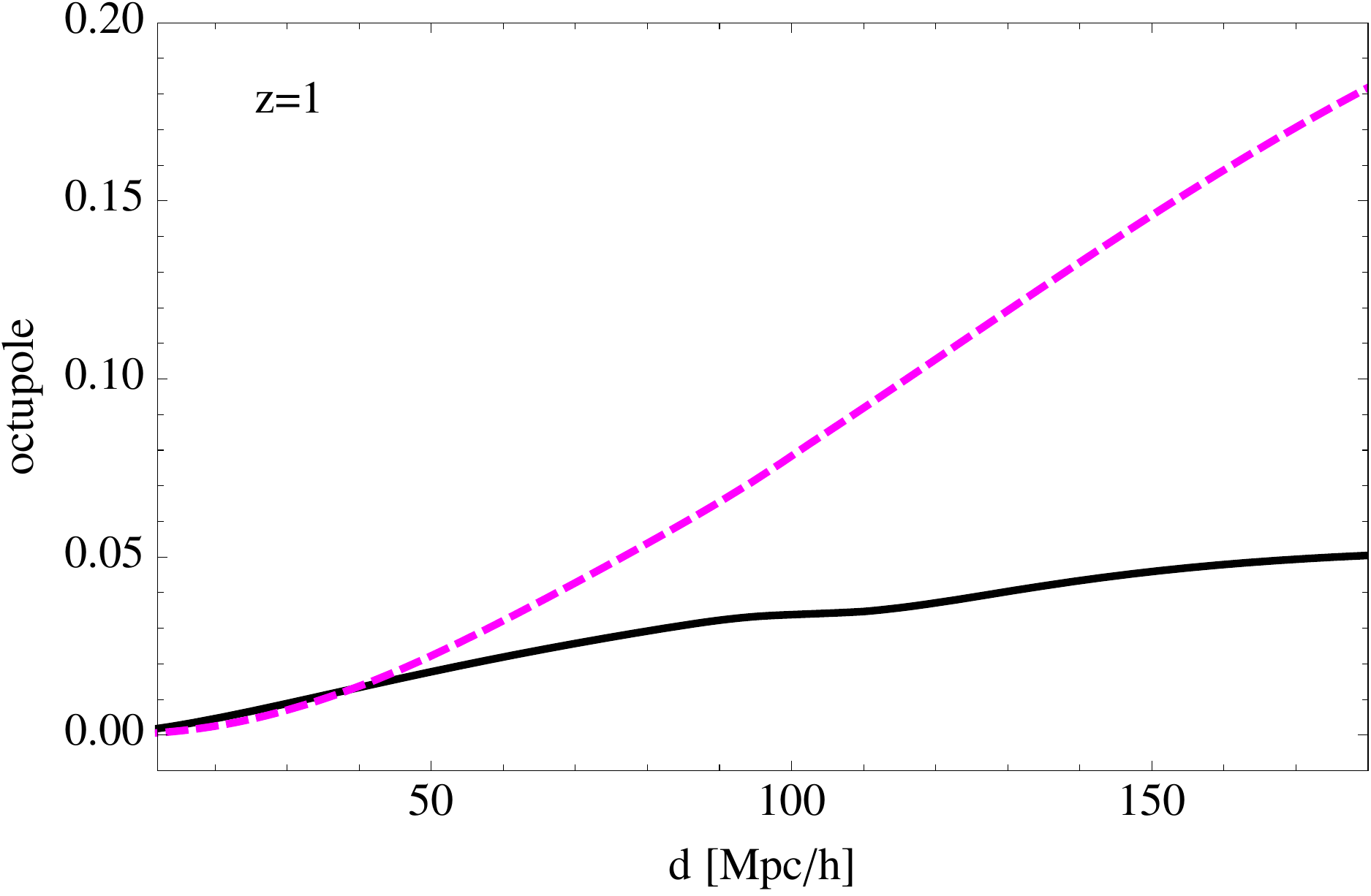}
\caption{\label{fig:lensing} {\it Left panels:} Amplitude of the Doppler magnification dipole (black solid line) and the gravitational lensing dipole (magenta dashed line) as a function of separation $d$, at $z=0.1$ and $z=1$. {\it Right panels:} Same for the octupole. In all plots, the dipole and octupole are multiplied by $d^2$.}
\end{figure*}

\begin{figure*}
\centering
\includegraphics[width=0.48\textwidth]{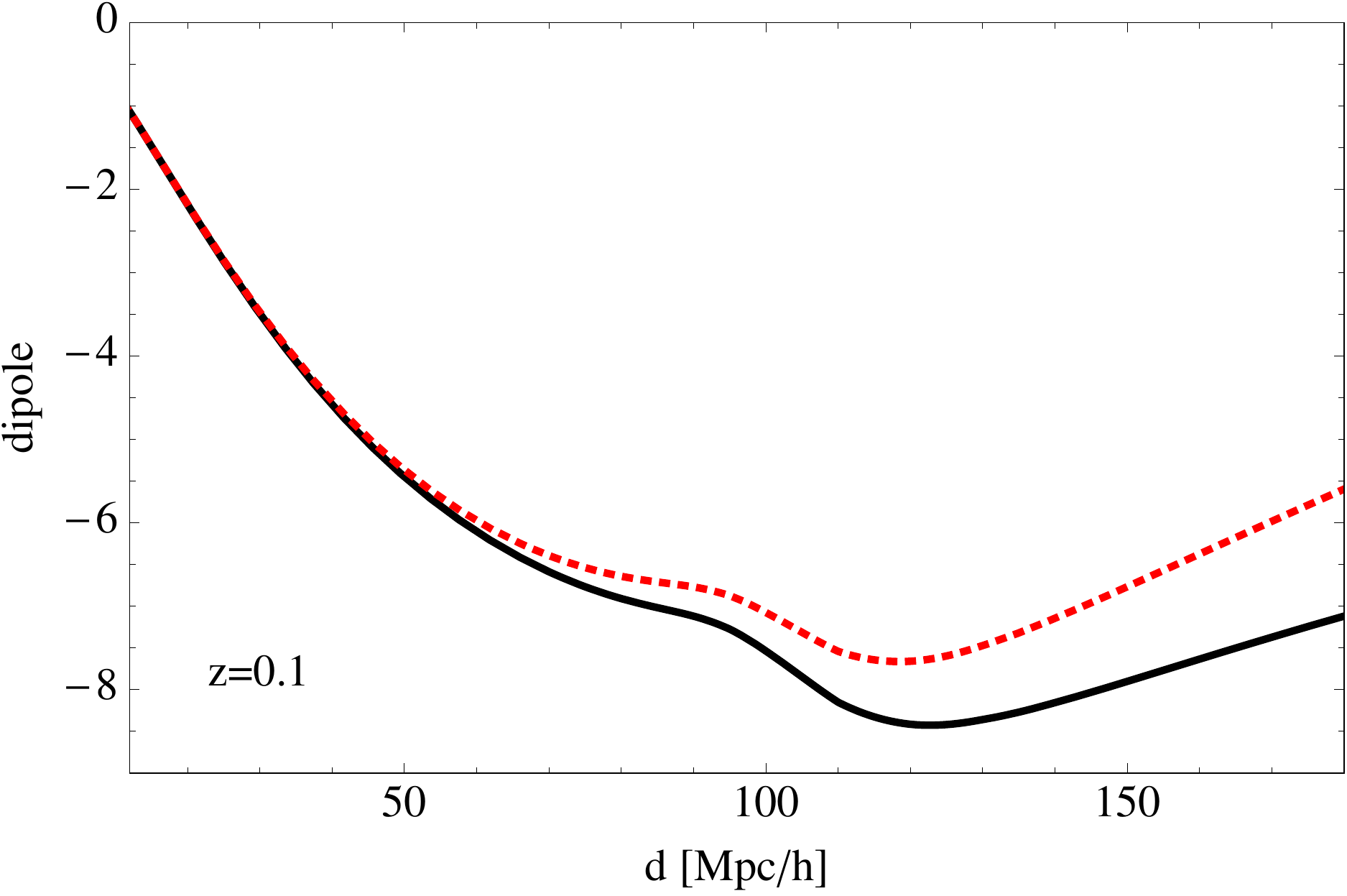}\hspace{0.5cm}\includegraphics[width=0.48\textwidth]{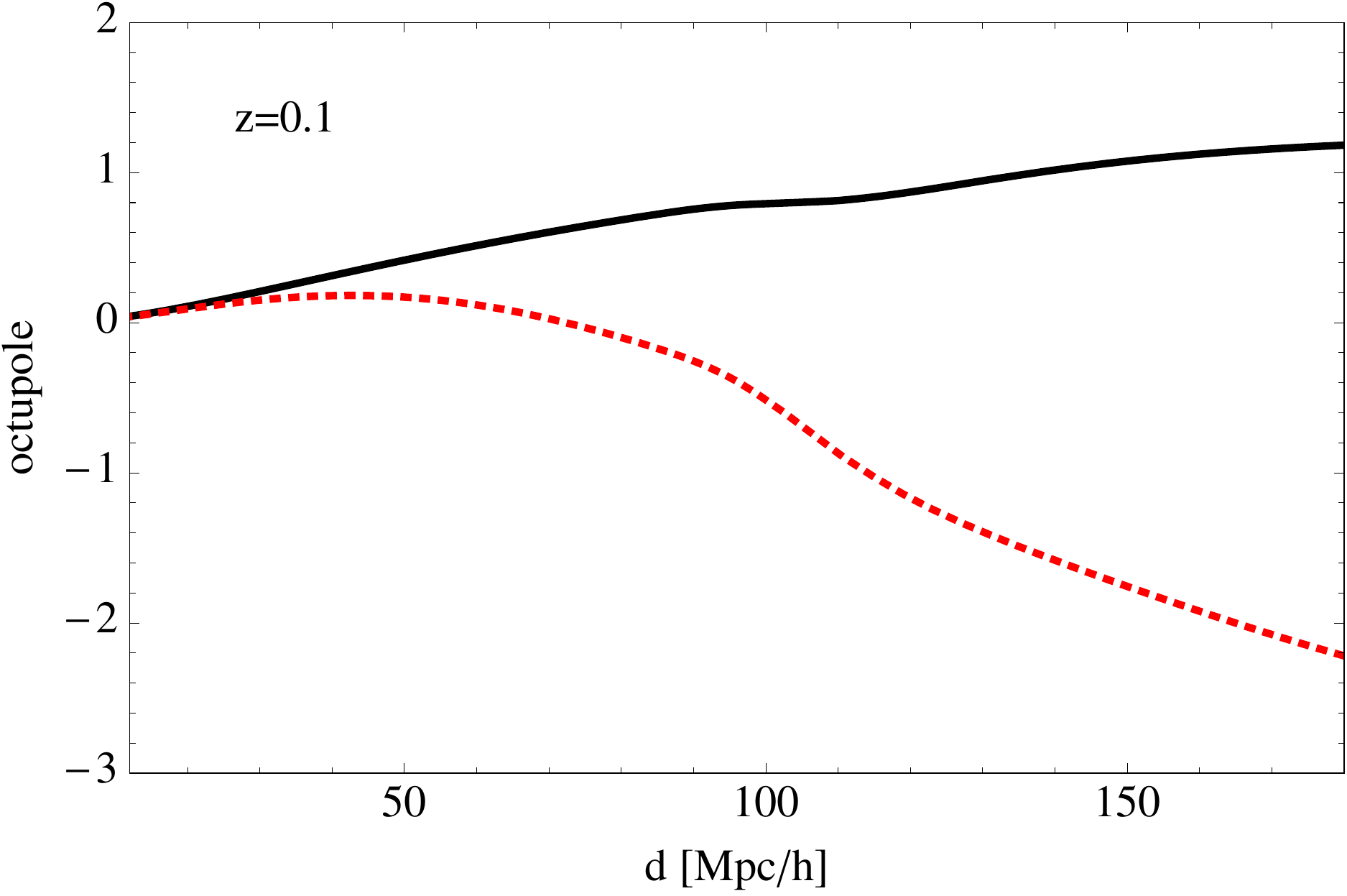}\\
\includegraphics[width=0.48\textwidth]{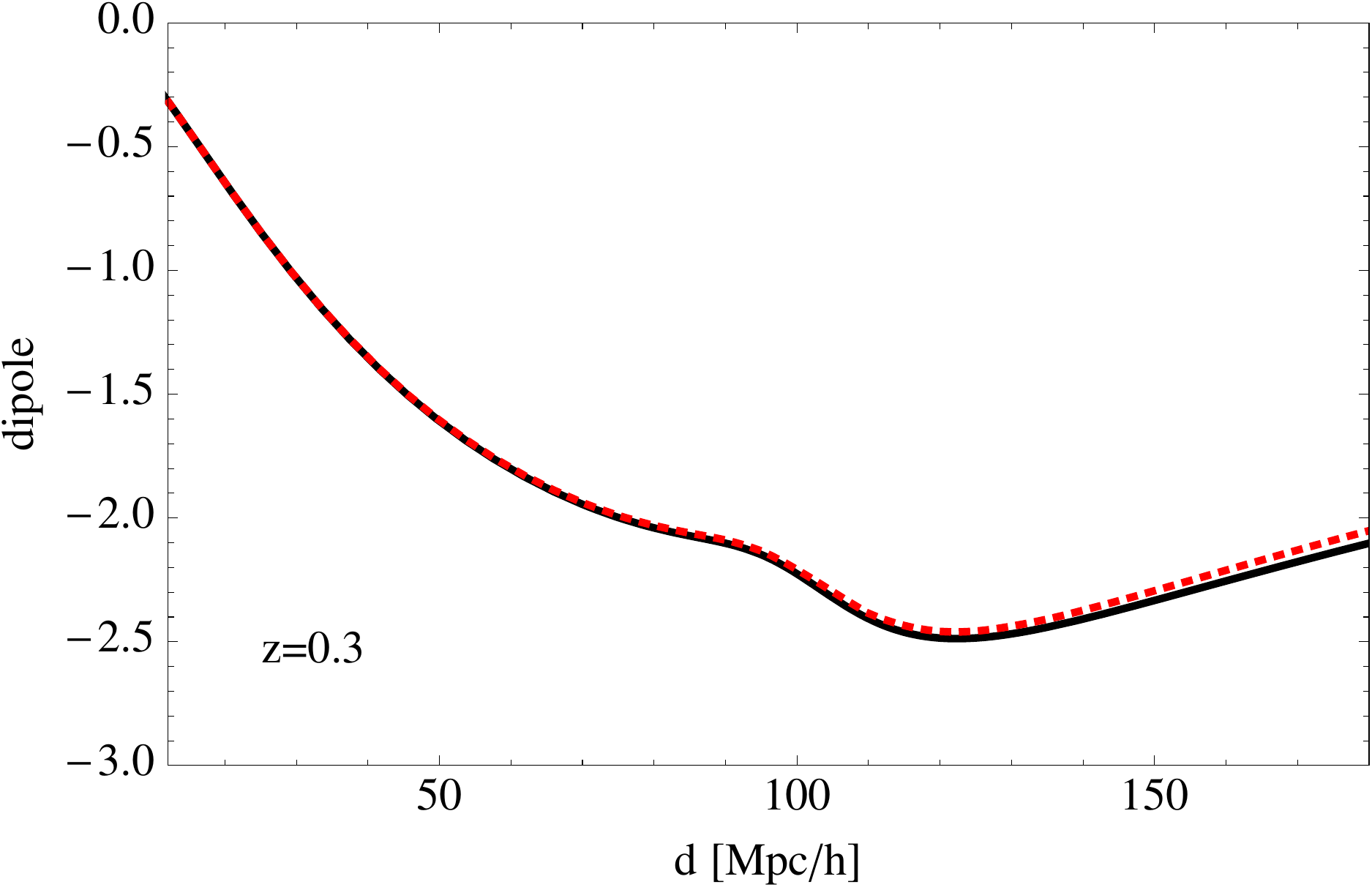}\hspace{0.5cm}\includegraphics[width=0.48\textwidth]{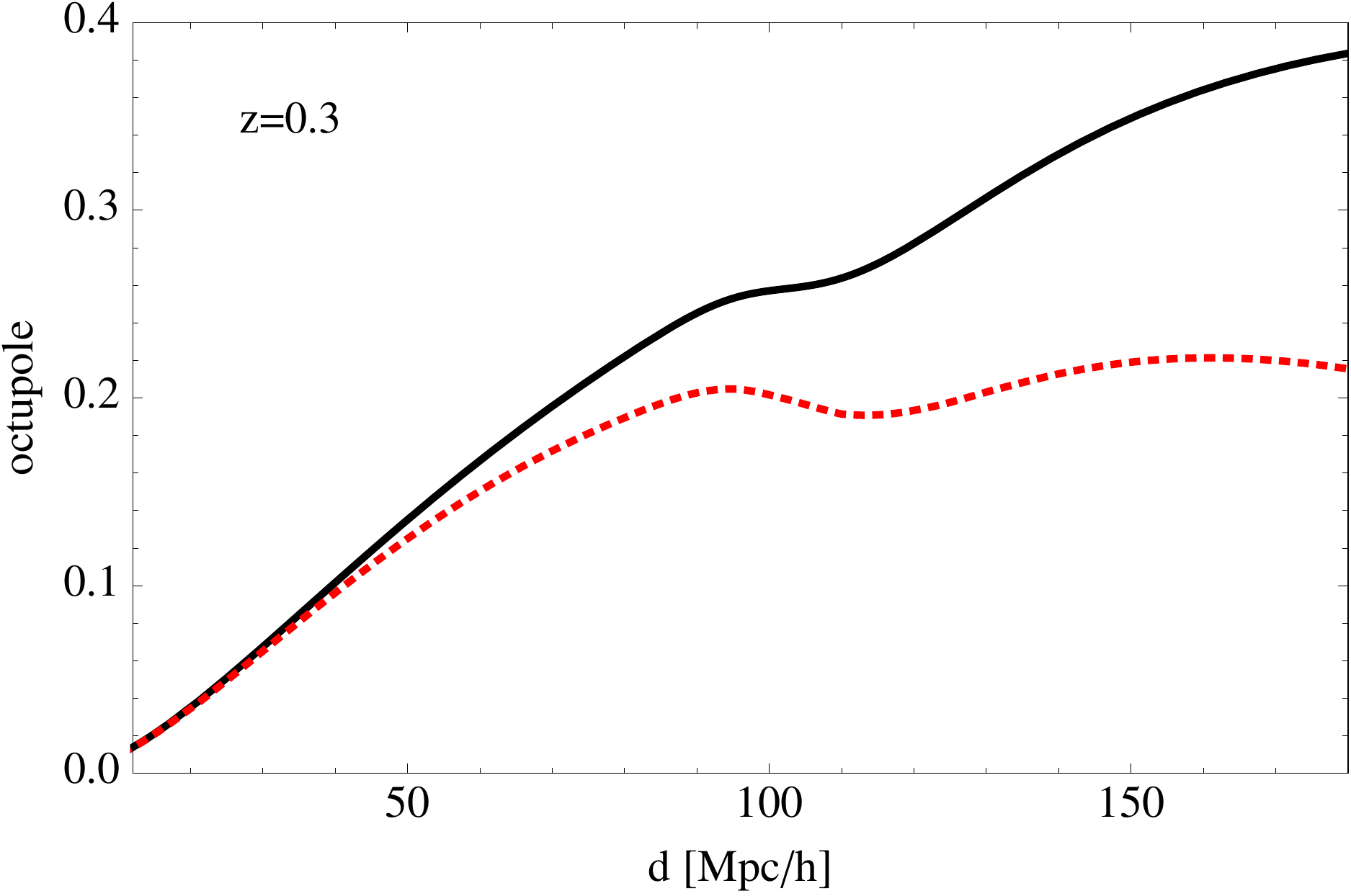}
\caption{\label{fig:LS}  {\it Left panels:} Amplitude of the Doppler magnification dipole as a function of separation $d$, at $z=0.1$ and $z=0.3$, calculated in the distant observer approximation Eq.~\eqref{meandip} (black solid line), and for the full sky (red dotted line). {\it Right panels:} Same for the octupole. In all plots the dipole and octupole are multiplied by $d^2$.}
\end{figure*}

Knowing the form of the Doppler magnification contribution we can construct an estimator to isolate it in the cross-correlation. From \eqref{xiv} we see immediately that in the distant observer approximation an obvious choice is to weight the correlation function by $P_1(\cos\beta)=\cos\beta$, and by $P_3(\cos\beta)$, and to integrate over $\beta$. In terms of discrete bins $i$ and $j$, we construct
\bea
\hat\xi_{\rm dip}(d)&=&\norm \sum_{ij} \Delta_i \kappa_j \cos\beta_{ij}\delta_K(d_{ij}-d)\, ,\label{estimatordip}\\
\hat\xi_{\rm oct}(d)&=&b_{\rm N} \sum_{ij} \Delta_i \kappa_j P_3(\cos\beta_{ij})\delta_K(d_{ij}-d)\, ,\label{estimatoroct}
\eea
where $\norm$ and $b_{\rm N}$ are normalisation factors, and $\delta_K$ denotes the Kronecker-$\delta$ function. Eqs.~\eqref{estimatordip} and~\eqref{estimatoroct} allow us to measure the amplitude of the dipole and of the octupole generated by the Doppler magnification. 
To determine the normalisation factors $\norm$ and $b_{\rm N}$ we take the continuous limit of~\eqref{estimatordip} and~\eqref{estimatoroct}. The derivation is presented in Appendix~\ref{app:mean}. We find
\be
\norm=\frac{3}{4\pi}\frac{\ell_p^5}{d^2V}\hspace{1cm}\mbox{and}\hspace{1cm} b_{\rm N}=\frac{7}{4\pi}\frac{\ell_p^5}{d^2V}\, ,
\ee
where  $\ell_p$ is the size of the cubic pixels in which we measure $\Delta$ and $\kappa$, and $V$ denotes the total volume of the survey (or the volume of the redshift bin in which we average the signal). Neglecting the lensing contribution, the mean of the estimators then becomes
\begin{align}
&\langle\hat{\xi}_{\rm dip}\rangle(d)\simeq\frac{\HH(z)}{\HH_0}f(z)\left(1-\frac{1}{\HH(z)r} \right)\left(b(z)+\frac{3f(z)}{5} \right)\nu_1(d)\, ,\label{meandip}\\
&\langle\hat{\xi}_{\rm oct}\rangle(d)\simeq-\frac{\HH(z)}{\HH_0}f(z)\left(1-\frac{1}{\HH(z)r} \right)\frac{2f(z)}{5}\nu_3(d)\, .\label{meanoct}
\end{align}

In Figure~\ref{fig:dipall}, we plot the dipole~\eqref{meandip} and the octupole~\eqref{meanoct} in a $\Lambda$CDM Universe with cosmological parameters $h=0.68, \,n_s=0.96, \,\Omega_{\rm cdm}=0.2548,\,\Omega_b=0.048$ and primordial amplitude of scalar perturbations $A=2.2 \times10^{-9}$ (corresponding to $\sigma_8=0.83$). We see that both the dipole and the octupole decrease quickly with redshift. As expected the amplitude of the dipole is negative: a galaxy situated behind an overdensity (with $\cos\beta=1$) is apparently demagnified by its peculiar motion and the correlation function is therefore negative. The octupole is generated by the correlation between the Doppler magnification and the redshift-space distortion experienced by the overdensity. We see that this contribution is positive and significantly smaller than the dipole. This difference in amplitude is due to the difference between $\nu_1(d)$ and $\nu_3(d)$ as well as to the different pre-factors of the dipole and the octupole. In particular the dipole is enhanced by the galaxy bias, which we choose here to evolve according to~\citet{1994ApJ...421L...1N, 1996ApJ...461L..65F, Tegmark:1998wm}
\be
b(z)=1+(b_i-1)\frac{D(z_i)}{D(z)}\,,
\ee
where $b_i$ is the initial value of the bias at redshift $z_i\simeq3$. We choose as an example $b_i$ such that $b=2$ at $z=0.5$.
Since the dipole is almost 10 times larger than the octupole it will be easier to detect.

As shown in Eq.~\eqref{xilens}, the gravitational lensing $\kappa_\g$ also generates an asymmetric contribution to the correlation function. This asymmetry will contribute to the estimator for the dipole and the octupole. In Figure~\ref{fig:lensing}, we compare the Doppler magnification multipoles with the gravitational lensing multipoles at two different redshifts. We see that at low redshift $z=0.1$, the gravitational lensing contribution to the dipole is completely negligible, less than a percent at all scales. As the redshift increases, the gravitational lensing contribution becomes more important. At redshift $z=0.3$, it remains very small, less than a few percent at all separations. It reaches 7\% at $z=0.5$ and $d=180$\,Mpc$/h$ and 21\% at $z=1$ and $d=180$\,Mpc$/h$. For redshifts $z\lsim 0.5$, then, the estimator~\eqref{estimatordip} provides a robust way of isolating Doppler magnification from gravitational lensing, and even at large redshift this estimator picks up the Doppler magnification predominantly. On the other hand, we find that the contamination from gravitational lensing to the octupole is more important: at $z=0.1$ and $d=180$\,Mpc$/h$ the lensing contribution is already 14\% of the Doppler contribution and at $z=0.5$ the lensing contribution dominates over the Doppler contribution. The octupole is therefore less efficient than the dipole for isolating the Doppler magnification.

\subsection{Validity of the distant observer approximation}
\label{sec:fullsky}

Eqs.~\eqref{meandip} and~\eqref{meanoct} are valid in the distant observer approximation, i.e. for $d\ll r$. At large separations, $d\sim r$, this approximation clearly breaks down and two types of correction come into play. First, there is a {\it wide-angle} correction: for large separations, the angle $\alpha$ differs from the angle $\beta$ (see Figure~\ref{fig:coordinate}), $\alpha=\beta - \theta\neq \beta$ for large $\theta$. Expanding Eqs.~\eqref{cosalpha} and~\eqref{sinalpha} in powers of $d/r$, we see that in Eq.~\eqref{xifull}, the difference between $\alpha$ and $\beta$ generates corrections of the order $d/r$ multiplied by even powers of $\cos\beta$ and corrections of the order $(d/r)^2$ multiplied by odd powers of $\cos\beta$. The second type of correction is due to {\it evolution}: these come from the fact that $r'\neq r$, and that the Hubble parameter $\HH$, growth rate $f$, and bias $b$ evolve with redshift. Using \eqref{rprime} and Taylor expanding $\HH$, $f$ and $b$ around $r$, we find that the evolution between $r$ and $r'$ in \eqref{xifull} also generates corrections of the order $d/r$ multiplied by even powers of $\cos\beta$ and corrections of the order $(d/r)^2$ multiplied by odd powers of $\cos\beta$. 

From this we understand that at large separations, Doppler magnification generates a monopole and a quadrupole, whose amplitudes are suppressed by $d/r$ with respect to the dipole and octupole. Furthermore, the distant observer expressions for the dipole and the octupole given in~\eqref{meandip} and~\eqref{meanoct} receive corrections proportional to $(d/r)^2$. 

In Figure~\ref{fig:LS}, we compare the distant observer expression for the dipole~\eqref{meandip} and the octupole~\eqref{meanoct}, with the full-sky result, obtained by inserting~\eqref{xifull} into~\eqref{estimatordip} and~\eqref{estimatoroct} and numerically integrating over the angle $\beta$. We see that at low redshift $z=0.1$, the corrections to the distant observer dipole are $\sim 7\%$ at $d=100$\,Mpc$/h$ and reach 27\% at $d=180$\,Mpc$/h$. At larger redshift, $z=0.3$, the distant observer dipole is a good approximation up to $d=180$\,Mpc$/h$ (where the correction is of order 2\%) and it becomes even more accurate at $z=0.5$ and $z=1$. This is simply due to the fact that the corrections to the dipole scale as $(d/r)^2$ and therefore decrease quickly as $r$ increases. From the right panel of Figure~\ref{fig:LS} we see that the wide-angle and evolution corrections to the octupole are significantly larger than for the dipole. This comes from the fact that at large scales there is a leaking of the dipole into the octupole: terms proportional to the bias and to $\nu_1(d)$ in Eq.~\eqref{xifull} contribute to the octupole at large separation, and since the dipole is 10 times larger than the octupole, these wide-angle corrections affect the octupole significantly.
In the following we forecast the signal-to-noise and cosmological constraints using the full-sky expression for the dipole and the octupole, since most of the constraining power comes from small redshifts, where the distant observer approximation quickly becomes inaccurate.

\section{Variance and signal-to-noise}
\label{sec:variance}

We now evaluate the signal-to-noise of the dipole and the octupole in various surveys. 

\subsection{Variance}

We start by calculating the variance of the dipole estimator~\eqref{meandip}.
We have
\begin{align}
&{\rm var}\Big(\hat \xi_{\rm dip}\Big)=\left\langle\left(\hat \xi_{\rm dip}\right)^2\right\rangle-\left\langle\hat \xi_{\rm dip}\right\rangle^2\label{vardip}\\
&= \frac{9\ell_p^{10}}{16\pi^2 V^2 d^2 d'^2} \sum_{ij}\sum_{ab}\Big[ \langle \Delta_i \kappa_j \Delta_a \kappa_b\rangle-\langle \Delta_i \kappa_j\rangle \langle\Delta_a \kappa_b\rangle \Big]\nonumber\\
&\quad\times \cos\beta_{ij}\cos\beta_{ab}\delta_K(d_{ij}-d)\delta_K(d_{ab}-d')\nonumber\\
&=\frac{9\ell_p^{10}}{16\pi^2 V^2 d^2 d'^2} \sum_{ij}\sum_{ab}\Big[ \langle \Delta_i \Delta_a\rangle \langle \kappa_j  \kappa_b\rangle+\langle \Delta_i \kappa_b\rangle \langle\Delta_a \kappa_j\rangle \Big]\nonumber\\
&\quad\times \cos\beta_{ij}\cos\beta_{ab}\delta_K(d_{ij}-d)\delta_K(d_{ab}-d')\, , \nonumber
\end{align}
where in the third equality we have used Wick's theorem, which is valid in the regime where the fields are Gaussian, i.e. when $\Delta$ and $\kappa$ are in the linear regime.
There are three types of contribution to the variance. First, $\langle \Delta_i \Delta_a\rangle$ contains a Poisson contribution and a cosmic variance contribution
\be
\label{DeltaDelta}
\langle \Delta_i \Delta_a\rangle=\frac{1}{\delta\bar n}\delta_{ia}+ C^\Delta_{ia}\, ,
\ee
where $\delta\bar n$ is the mean number of galaxies per pixel. In the distant observer approximation the cosmic variance $C^\Delta$ is given by
\begin{align}
C^\Delta_{ia}=&\frac{1}{(2\pi)^3}\int d^3 \bk\, e^{i\bk(\bx_a-\bx_i)}P(k,z)\Bigg[b^2+\frac{2bf}{3}+\frac{f^2}{5} \nonumber\\
&+\left(\frac{4bf}{3}+\frac{4f^2}{7} \right)P_2(\hat\bk\cdot\bn)+\frac{8f^2}{35} P_4(\hat\bk\cdot\bn)\Bigg]\, .
\end{align}
The relative importance of the Poisson noise and the cosmic variance depends on the characteristics of the survey and on the separation $d_{ia}$.

Second, $\langle \kappa_j  \kappa_b\rangle$ contains an intrinsic error on the measurement of the galaxy's size and a cosmic variance contribution
\be
\langle \kappa_j  \kappa_b\rangle=\sigma^2_\kappa\delta_{jb}+ C^\kappa_{jb}\, .
\ee
The amplitude of the intrinsic error $\sigma_\kappa$ depends on the type of galaxy, as well as on the resolution of the instrument~\citep{2012ApJ...744L..22S, Casaponsa:2012tq, Heavens:2013gol, Alsing:2014fya}. In the following we consider two values: an optimistic value $\sigma_\kappa=0.3$ and a more pessimistic value $\sigma_\kappa=0.8$. The cosmic variance $C^\kappa$ is at most of the order $10^{-4}$ in the range of redshifts we are interested in and it can therefore be safely neglected with respect to the intrinsic contribution. 

Finally, $\langle \Delta_i \kappa_j\rangle$ only contains a contribution from cosmic variance. This contribution is nothing other than our signal, which we found to be on the order of $10^{-2}$ at most, as can be seen from Figure~\ref{fig:dipall} (where the amplitude should be divided by $d^2$). We therefore see that the second contribution in \eqref{vardip} is always subdominant with respect to the first contribution and we neglect it in the following. 

\begin{figure*}
\centering
\includegraphics[width=0.48\textwidth]{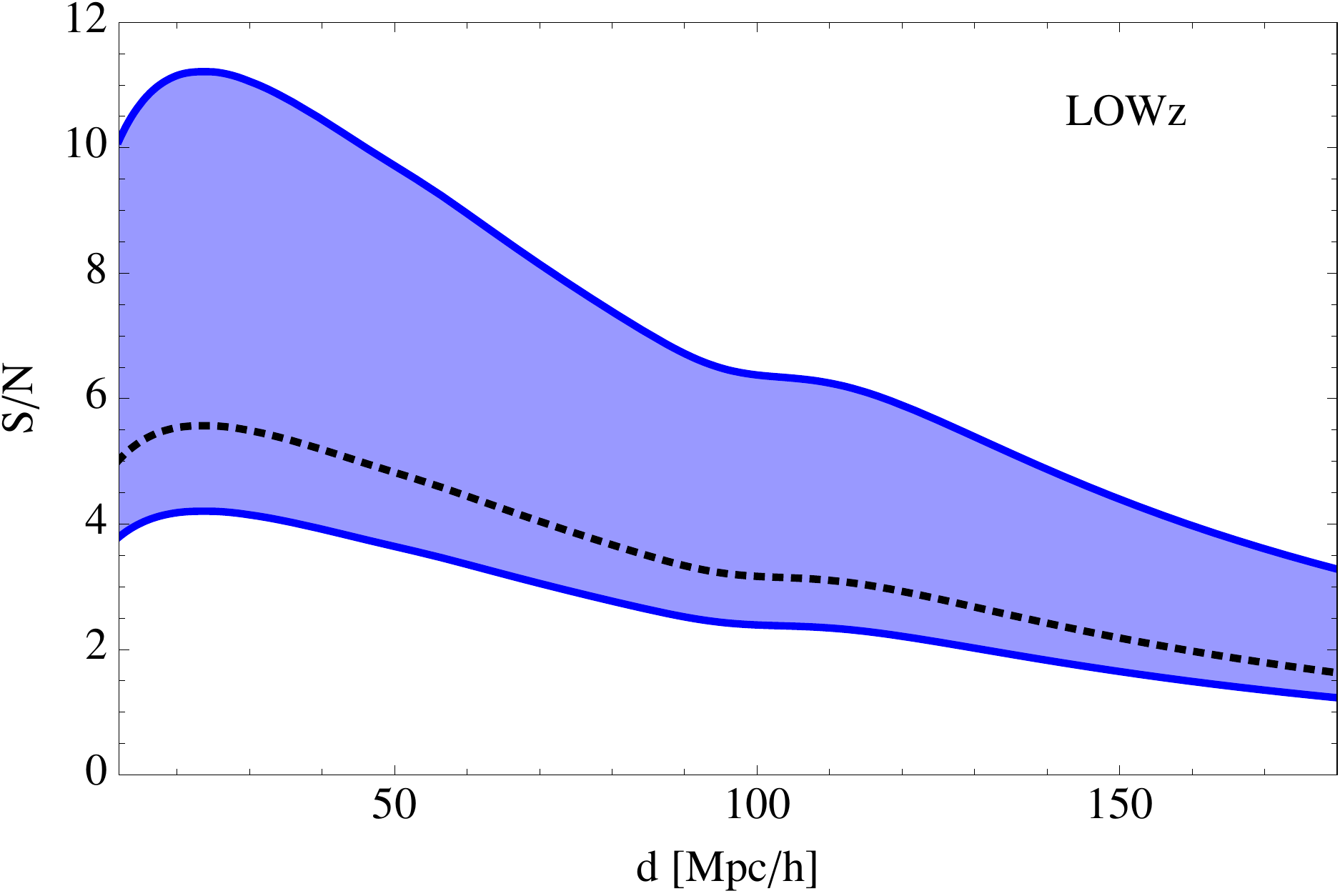}\hspace{0.5cm}
\includegraphics[width=0.48\textwidth]{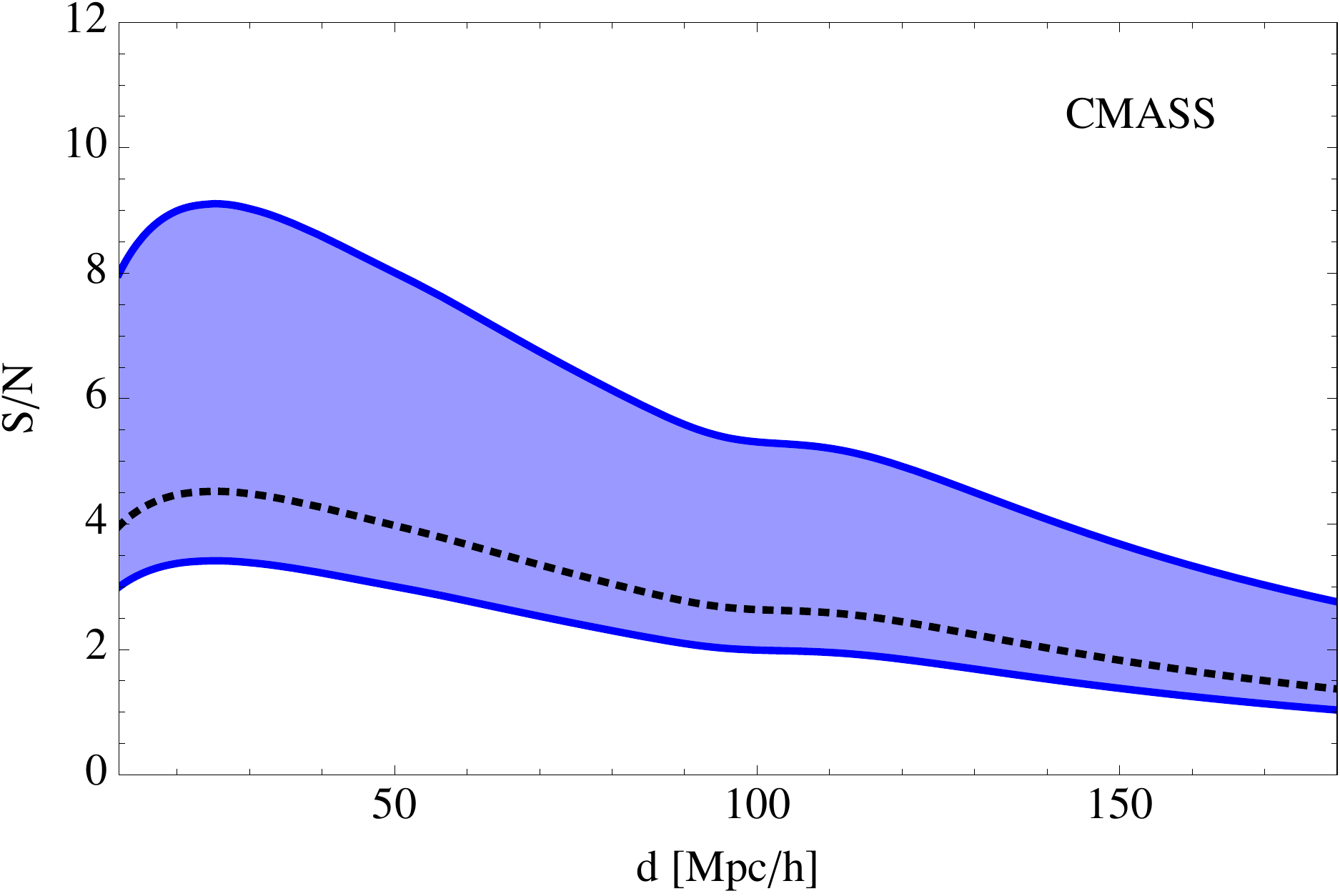}
\caption{\label{fig:SN_opt} Signal-to-noise for the dipole in LOWz and CMASS, plotted as a function of separation. The higher bound corresponds to an intrinsic error on the size measurement of $\sigma_\kappa=0.3$, and the lower bound of $\sigma_\kappa=0.8$. The dotted black line corresponds to a mixed sample with 50\% of galaxies with $\sigma_\kappa=0.3$ and 50\% with $\sigma_\kappa=0.8$.}
\end{figure*}

We then obtain
\begin{align}
{\rm var}\Big(\hat \xi_{\rm dip}\Big)&\simeq\frac{9\ell_p^{10}\sigma_\kappa^2}{16\pi^2 V^2 d^2 d'^2}\label{variance}\\
&\times \Bigg[\frac{1}{\delta\bar n}\sum_{ij}\cos^2\beta_{ij}
\delta_K(d_{ij}-d)\delta_K(d-d')\nonumber\\
&+\sum_{ija}C^\Delta_{ia}\cos\beta_{ij}\cos\beta_{aj}\delta_K(d_{ij}-d)\delta_K(d_{aj}-d') \Bigg]\, . \nonumber
\end{align}
The first term in~\eqref{variance} can easily be calculated in the continuous limit by fixing the position of the pixel $i$ and integrating over $j$. We obtain
\be
{\rm var}_1\Big(\hat \xi_{\rm dip}\Big)=\frac{3}{4\pi}\frac{\ell_p^2}{d^2}\frac{\sigma^2_\kappa}{N_{\rm tot}}\delta_K(d-d')\, , \label{var1}
\ee
where $N_{\rm tot}$ is the total number of galaxies in the (number count) survey,
\be
N_{\rm tot}=\frac{\delta\bar n}{\ell_p^3}V\, .
\ee
The second term in~\eqref{variance} contains a sum over 3 pixels. We calculate this term using the method presented in~\citet{Hall:2016bmm}. We obtain (see Appendix~\ref{app:variance} for more detail)
\begin{align}
{\rm var}_2\Big(\hat \xi_{\rm dip}\Big)=&\frac{9}{2\pi^2}\frac{\ell_p^3}{V}\sigma^2_\kappa\left(\frac{b^2}{3}+\frac{2bf}{5}+\frac{f^2}{7} \right)\label{var2}\\
&\times\int dk k^2 P(k,z)j_1(kd)j_1(kd')\, . \nonumber
\end{align}
The first contribution~\eqref{var1} is diagonal, i.e. it vanishes for $d\neq d'$. The second contribution~\eqref{var2}, on the other hand, is non-diagonal and induces correlations between different pixel separations. The ratio between the first (Poisson) and second (cosmic variance) contributions is governed by
\be
\frac{{\rm var}_1}{\rm{var}_2}\propto \frac{1}{\bar n d^2 \ell_p}\, ,
\ee
where $\bar n$ denotes the mean number density. As expected, cosmic variance becomes more and more important at large separation. We also see that in surveys with high number density, the cosmic variance contribution dominates over the Poisson contribution.

A similar derivation can be made for the variance of the octupole. We find
\be
{\rm var}_1\Big(\hat \xi_{\rm oct}\Big)=\frac{7}{4\pi}\frac{\ell_p^2}{d^2}\frac{\sigma^2_\kappa}{N_{\rm tot}}\delta_K(d-d')\, , \label{var1oct}
\ee
and
\begin{align}
{\rm var}_2\Big(\hat \xi_{\rm oct}\Big)=&\frac{49}{2\pi^2}\frac{\ell_p^3}{V}\sigma^2_\kappa\left(\frac{b^2}{7}+\frac{46bf}{315}+\frac{13f^2}{231} \right)\label{var2oct}\\
&\times\int dk k^2 P(k,z)j_3(kd)j_3(kd')\,. \nonumber
\end{align}

Eqs.~\eqref{var1},~\eqref{var2},~\eqref{var1oct} and~\eqref{var2oct} assume that the sizes of all galaxies in the survey are measured with the same error, $\sigma_\kappa$. In reality, surveys are composed of various types of galaxies which may have different size errors. For example, as discussed in~\cite{Alsing:2014fya}, the sizes of late-type (spiral) galaxies tend to be better measured than for early-type (elliptical) galaxies. In Appendix~\ref{app:kappamixed} we show that, in this case, the variance keeps the same form as previously, but with $\sigma_\kappa$ replaced by an effective mixed error,
\be
\label{sigma_mixed}
\sigma^{\rm mixed}_\kappa=\sqrt{\frac{N^{\rm E}_{\rm tot}}{N_{\rm tot}}\big(\sigma^{\rm E}_\kappa\big)^2
+\frac{N^{\rm S}_{\rm tot}}{N_{\rm tot}}\big(\sigma^{\rm S}_\kappa\big)^2}\, ,
\ee
where $N^{\rm E}_{\rm tot}$ and $N^{\rm S}_{\rm tot}$ respectively denote the number of elliptical and spiral galaxies, and $\sigma^{\rm E}_\kappa$ and $\sigma^{\rm S}_\kappa$ are their associated size uncertainties. As an example, if we have a survey consisting of 50\% elliptical galaxies with $\sigma^{\rm E}_\kappa=0.3$ and 50\% spiral galaxies with $\sigma^{\rm S}_\kappa=0.8$, we obtain $\sigma^{\rm mixed}_\kappa=0.6$. Note that, as discussed in~\cite{Alsing:2014fya}, these numbers are likely to change, since new techniques may be developed in future to reduce the error on the size measurement of both elliptical and spiral galaxies.

Finally, in Appendix~\ref{app:thick} we also calculate similar expressions for the mean and variance of the dipole and the octupole for the case where the signal is averaged over a wide range of separation $d_{\rm min}\leq d\leq d_{\rm max}$.

\subsection{Signal-to-noise}

We calculate the signal-to-noise of the dipole and octupole in various surveys. We assume a pixel size of $\ell_p=4\,{\rm Mpc}/h$ and calculate the signal-to-noise for fixed separation $d$ between the pixels, where $d$ is a multiple of the pixel size.  
We have 
\bea
\frac{S}{N}(d)&=&\frac{\langle\hat{\xi}_{X}\rangle(d)}{\sqrt{{\rm var}_X(d)}}\, ,
\eea
where the mean and the variance are given by Eqs.~\eqref{meandip}, \eqref{var1} and \eqref{var2} when $X=$ dipole and by Eqs.~\eqref{meanoct}, \eqref{var1oct} and \eqref{var2oct} when $X=$ octupole. Note that here we calculate the signal-to-noise of the Doppler magnification only, neglecting the gravitational lensing contribution. As seen in Section~\ref{sec:fullsky} this is an excellent approximation for the dipole below $z=0.5$ but it is not a good approximation for the octupole, even at low redshift. However, as discussed at the end of Section~\ref{sec:fisher}, the signal-to-noise of the octupole is too small to impact the constraints on cosmological parameters and so we do not include it in our forecasts.

\begin{figure*}
\centering
\includegraphics[width=0.48\textwidth]{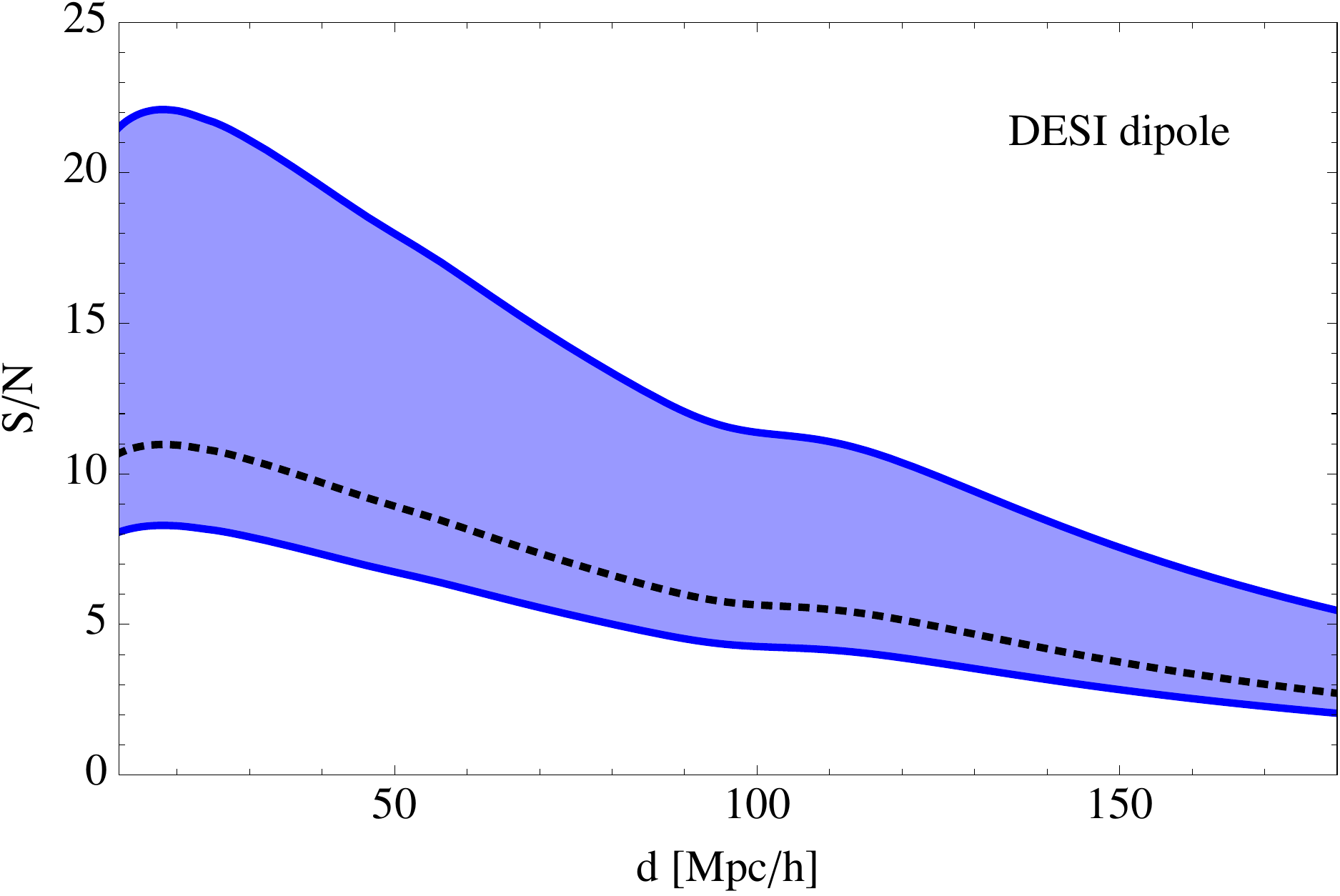}\hspace{0.5cm}\includegraphics[width=0.48\textwidth]{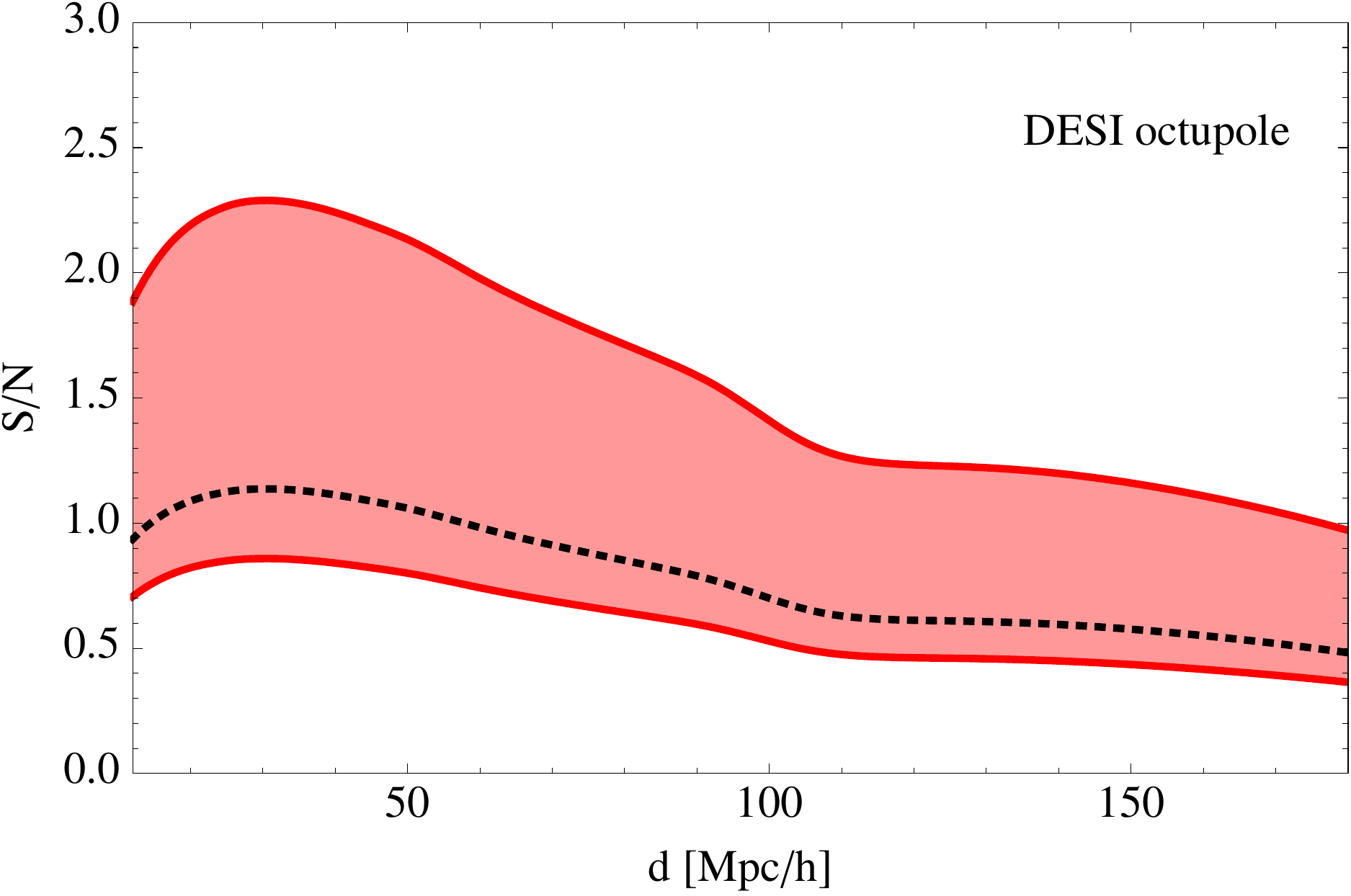}
\caption{\label{fig:SN_DESI} Predicted signal-to-noise for the dipole and octupole in the DESI Bright Galaxy sample, plotted as a function of separation. The higher bound corresponds to an intrinsic error on the size measurement of $\sigma_\kappa=0.3$, and the lower bound of $\sigma_\kappa=0.8$. The dotted black line corresponds to a mixed sample with 50\% of galaxies with $\sigma_\kappa=0.3$ and 50\% with $\sigma_\kappa=0.8$.}
\end{figure*}

\begin{figure*}
\centering
\includegraphics[width=0.48\textwidth]{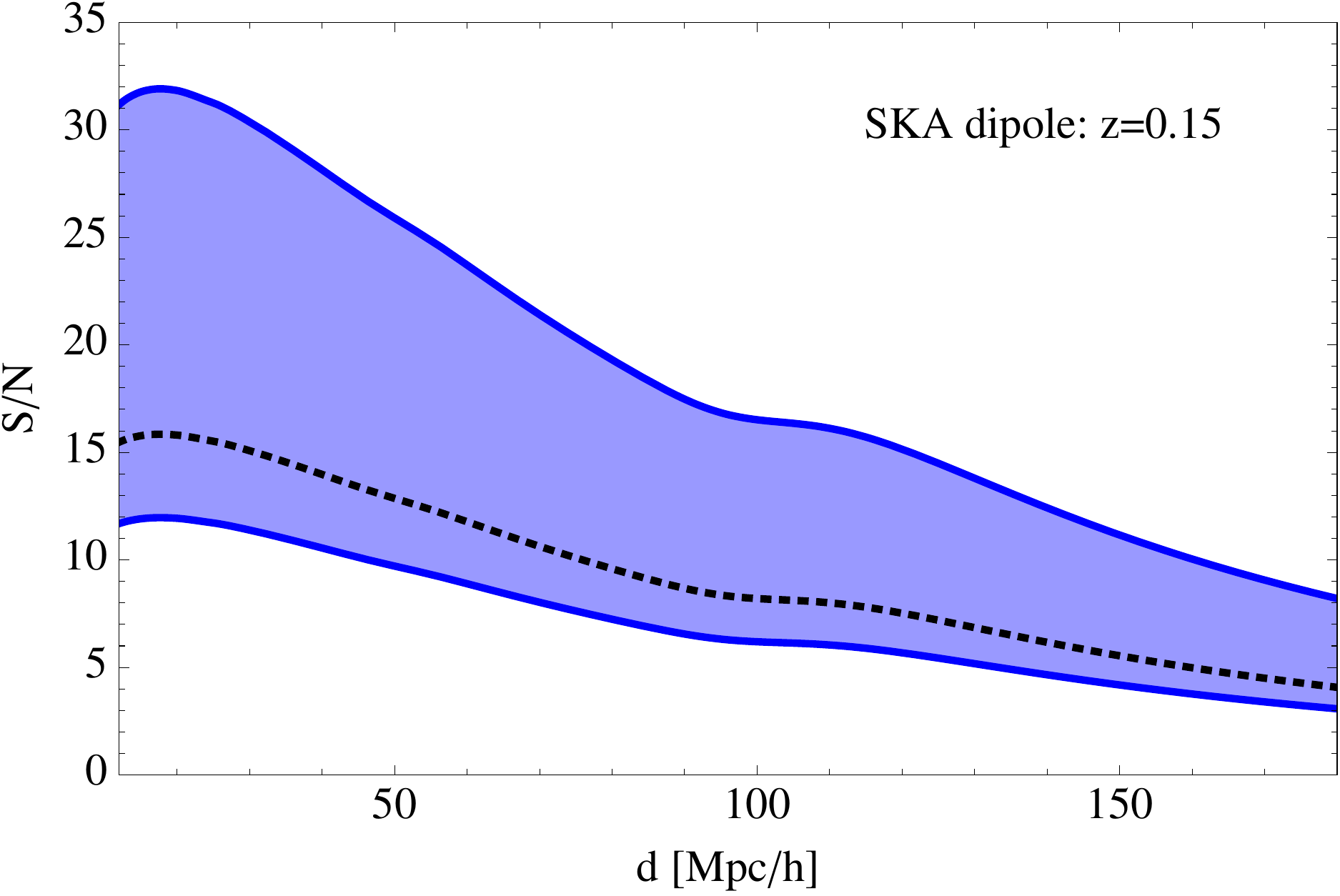}\hspace{0.5cm}\includegraphics[width=0.48\textwidth]{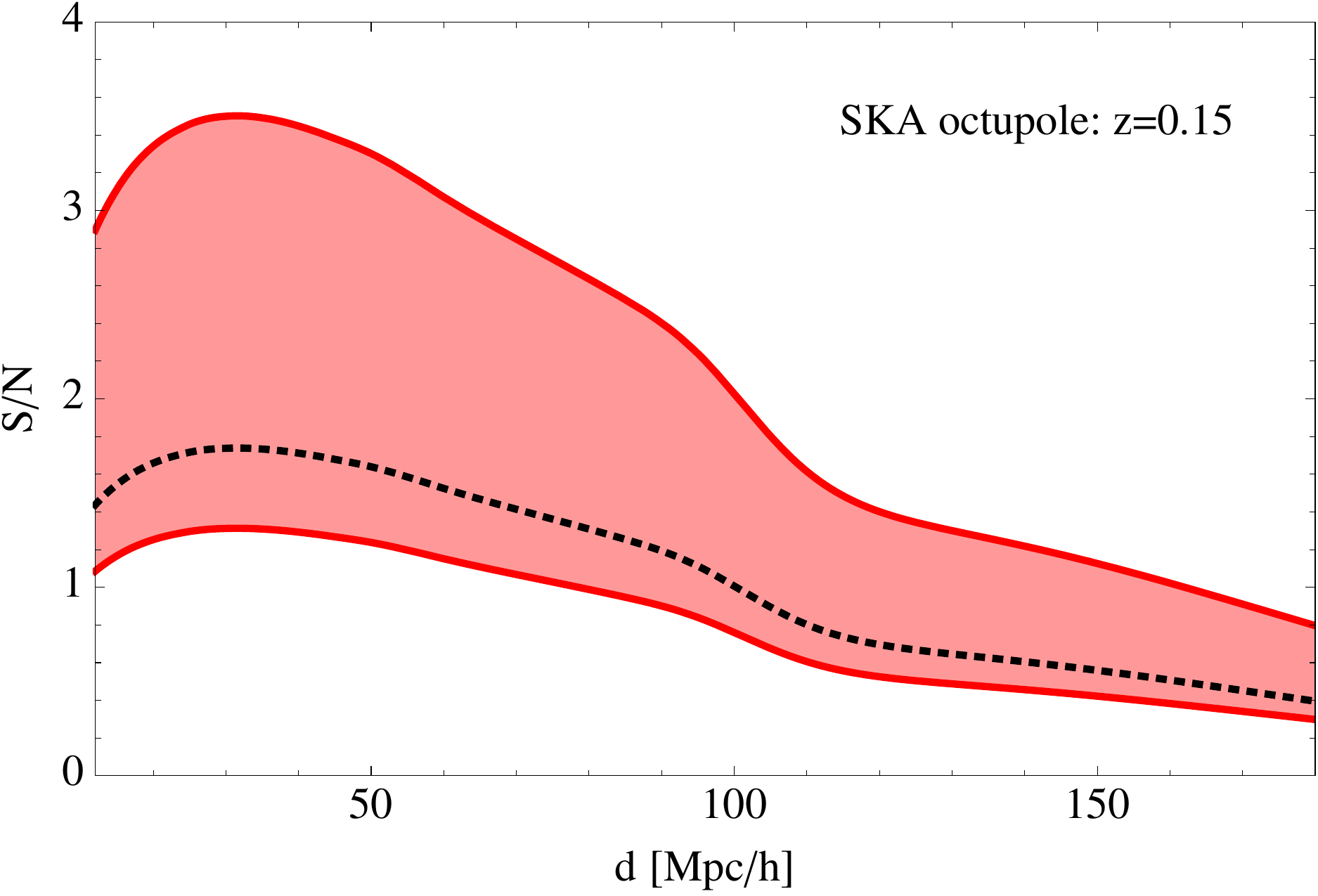}\\
\includegraphics[width=0.48\textwidth]{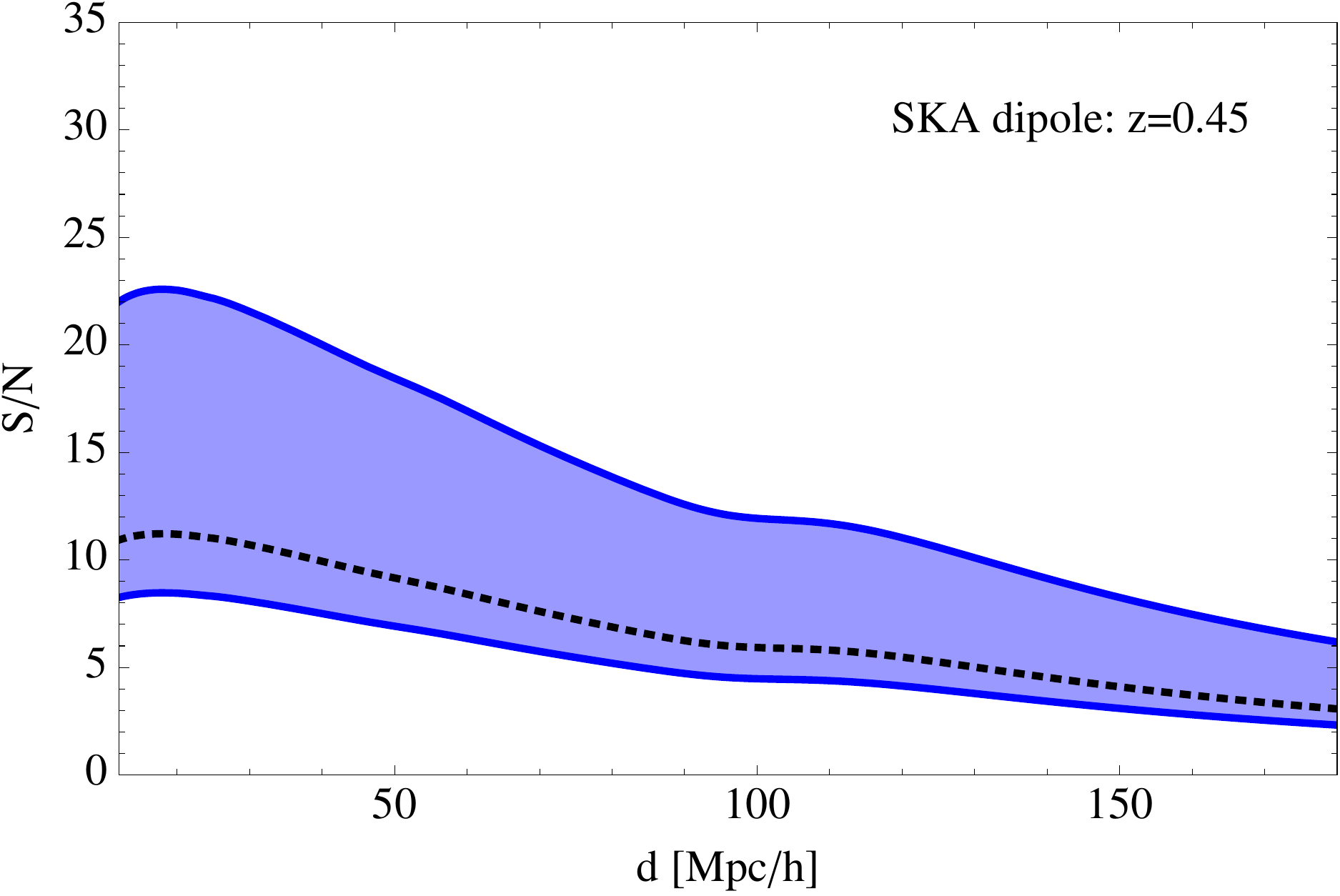}\hspace{0.5cm}\includegraphics[width=0.48\textwidth]{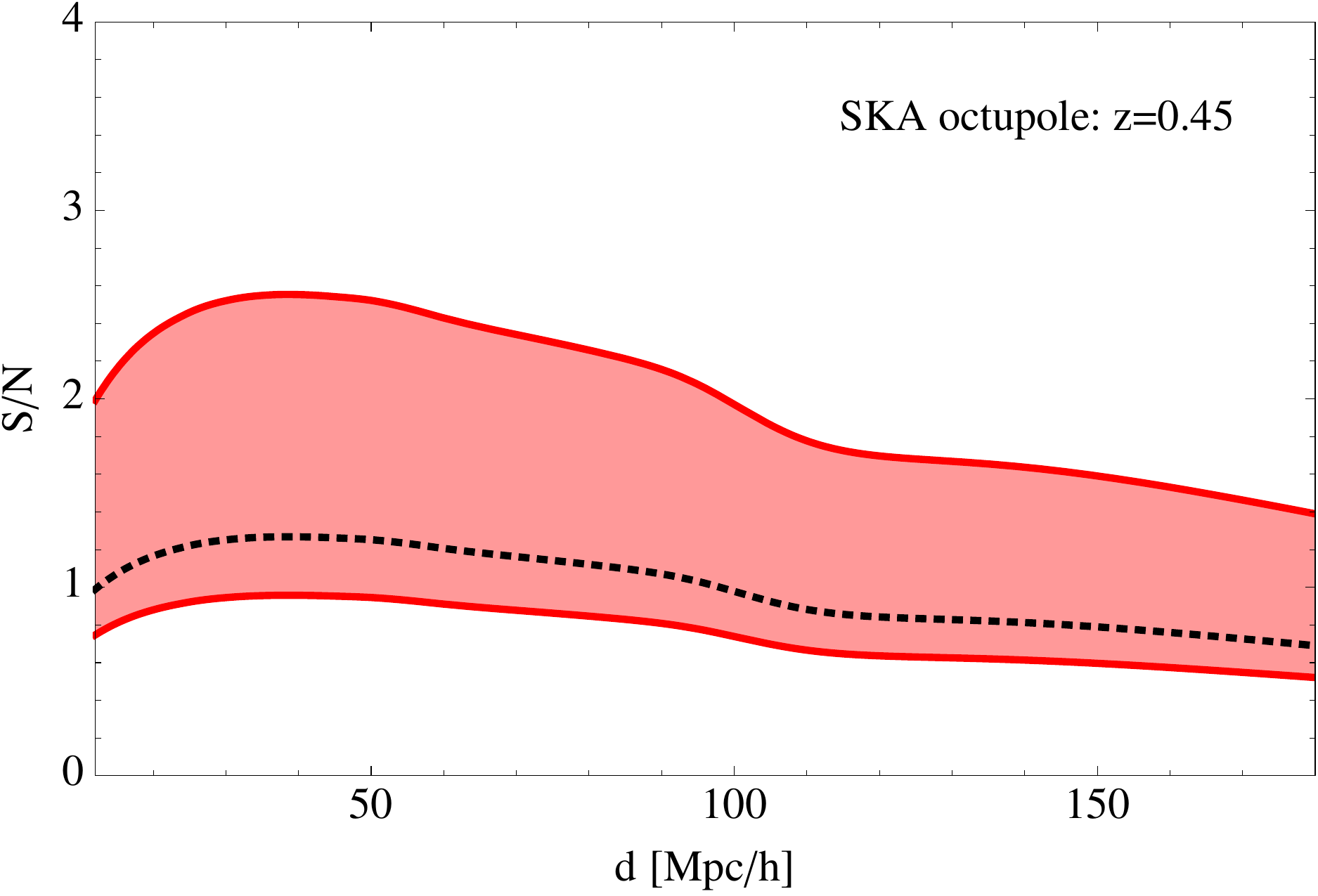}
\caption{\label{fig:SN_SKA} Signal-to-noise for the dipole and the octupole in the SKA Phase 2 survey, plotted as a function of separation. Here, we plot the signal-to-noise calculated in two thin redshift bins: $0.1<z<0.2$ and $0.4<z<0.5$. In each plot the higher bound corresponds to an intrinsic error on the size measurement of $\sigma_\kappa=0.3$, and the lower bound of $\sigma_\kappa=0.8$. The dotted black line corresponds to a mixed sample with 50\% of galaxies with $\sigma_\kappa=0.3$ and 50\% with $\sigma_\kappa=0.8$.}
\end{figure*}

We first calculate the signal-to-noise for current optical surveys. We consider three samples: the main sample of SDSS galaxies at $z\leq 0.2$, the LOWz sample and the CMASS sample. The volume, number density and mean bias are taken from~\citet{Percival:2006gt, Anderson:2013zyy, Gaztanaga:2015jrs}. We assume that for each galaxy in those samples we have a measurement of the size and magnitude from which we can infer the convergence using the estimator described in~\citet{2012ApJ...744L..22S, Casaponsa:2012tq, Heavens:2013gol, Alsing:2014fya}. The signal-to-noise for the dipole in LOWz and CMASS is plotted in Figure~\ref{fig:SN_opt}. The higher bound corresponds to an intrinsic error on the size measurement of $\sigma_\kappa=0.3$ and the lower bound of $\sigma_\kappa=0.8$. The dotted black line corresponds to a mixed sample with 50\% elliptical galaxies with $\sigma^{\rm E}_\kappa=0.3$ and 50\% spiral galaxies with $\sigma^{\rm S}_\kappa=0.8$. Naively one would expect the signal-to-noise of this mixed sample to be in the middle of the coloured region. However as shown in~\eqref{sigma_mixed} the uncertainties on $\kappa$ add in quadrature leading to $\sigma^{\rm mixed}_\kappa=0.6$ and not 0.5, which explains why the dotted line is closer to the lower boundary. The signal-to-noise is high enough to allow a detection of the Doppler magnification dipole in these two samples. In the main sample of SDSS, the signal-to-noise of the dipole reaches 2-6 (corresponding to $\sigma_\kappa=0.8$ and $\sigma_\kappa=0.3$ respectively) at low separation $d\lsim 50$\,Mpc/$h$. The octupole on the other hand has a signal-to-noise significantly smaller than one and can therefore not be detected in these samples.

The cumulative signal-to-noise over all separations can be calculated by accounting for the fact that the signal at different separations is correlated  
\be
\left(\frac{S}{N}\right)^2_{\rm cum}=\sum_{ab}\langle\hat{\xi}_{X}\rangle(d_a){\rm var}_X^{-1}(d_a, d_b)
\langle\hat{\xi}_{X}\rangle(d_b)\, .
\ee 
We find a cumulative signal-to-noise for the range of separation $12\leq d \leq 180$\,Mpc/$h$ of $3.8-10$  in the SDSS main sample, $8.4-23$ in LOWz and $7.3-20$ in CMASS. Assuming that the three samples are uncorrelated, we reach a total signal-to-noise of $12-31$. A robust detection of the Doppler magnification dipole should therefore be possible with current optical surveys. 

We then forecast the signal-to-noise for the future Dark Energy Spectroscopic Instrument (DESI)~\citep{Levi:2013gra}, along with imaging for the galaxies. The Bright Galaxy DESI survey~\citep{2015AAS...22533610C} will observe 10 million galaxies over 14,000 square degrees at redshift $z\leq 0.3$. To calculate the signal-to-noise in that range, we split the sample into three thin redshift bins: $0.05<z<0.1$, $0.1<z<0.2$ and $0.2<z<0.3$ that we assume to be uncorrelated.\footnote{We restrict the lower redshift to $z_{\rm min}=0.05$ in order to reduce the impact of local non-linear effects.} We assume a mean bias of $b=1.17$ over the whole sample, similar to the one of the main SDSS sample~\citep{Percival:2006gt}. In Figure~\ref{fig:SN_DESI} we show the signal-to-noise as a function of separation for the dipole and the octupole. The cumulative signal-to-noise over all separations is $14-37$ for the dipole and $1.9-5$ for the octupole. The dipole should therefore be robustly detected. The octupole will be difficult to see if $\sigma_\kappa$ is as large as 0.8 but may be just about detectable with a smaller size error. 

Finally we calculate the signal-to-noise for the SKA. In its second phase of operation the SKA HI (21cm) galaxy survey will detect galaxies spectroscopically  from redshift 0 to 2 over $\sim$30,000 square degrees. We forecast the signal-to-noise of the dipole and octupole from redshift $0.1$ to $0.5$, using the specifications from~\citet{Bull:2015lja} (see table 3). In Figure~\ref{fig:SN_SKA} we show the signal-to-noise for the dipole and the octupole in the lowest and highest redshift bins: $0.1<z<0.2$ and $0.4<z<0.5$.  Even though the volume (and consequently the number of galaxies) increases with redshift, the signal-to-noise decreases slightly since the signal is significantly larger at small redshift due to the $1/r$  dependence of the dipole and octupole amplitude, as seen from Eq.~\eqref{xifull}. 

The cumulative signal-to-noise over the range of separation $12\leq d \leq 180$\,Mpc/$h$, combining redshifts $0.1\leq z\leq 0.5$ (assuming that the redshift bins are uncorrelated), is $35-93$ for the dipole and $5.1-14$ for the octupole. The SKA should therefore allow us to robustly detect both the Doppler magnification dipole and octupole. Note that by going to higher redshifts we can slightly increase the signal-to-noise. For example the cumulative signal-to-noise of the dipole including data up to $z=0.8$ increases to $40-106$. Above redshift 0.5, the contribution to the dipole from gravitational lensing is no longer negligible however, and isolating the Doppler contribution therefore becomes more difficult. Both must be modelled together or measured.

\section{Forecasts}
\label{sec:fisher}

\begin{figure*}
\centering
\includegraphics[width=0.48\textwidth]{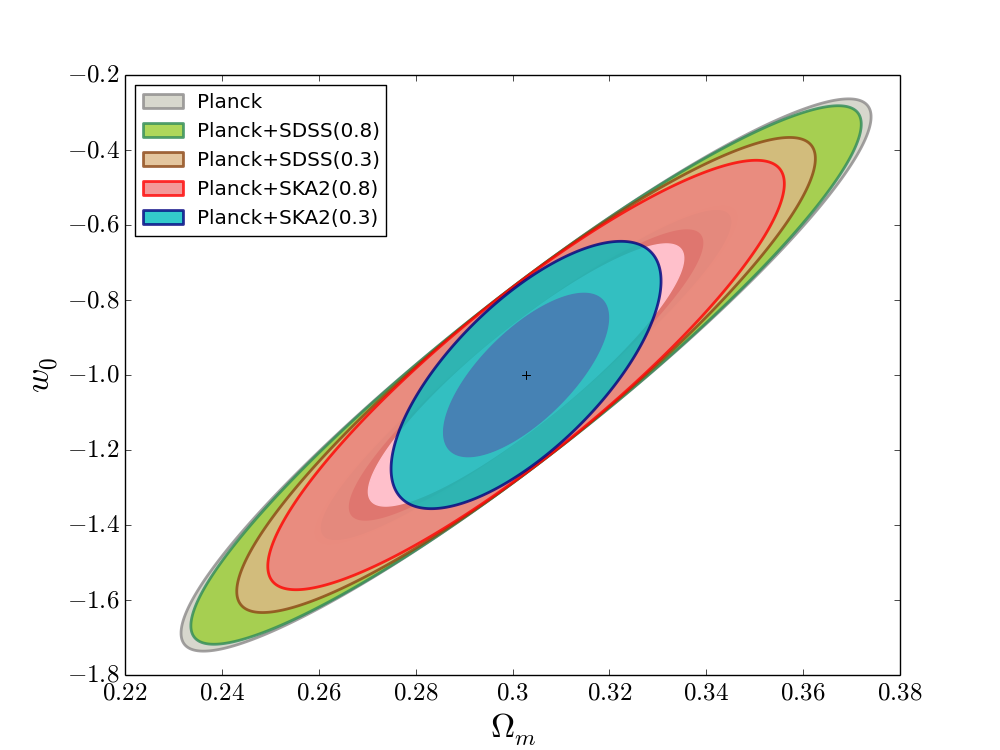}\hspace{0.5cm}
\includegraphics[width=0.48\textwidth]{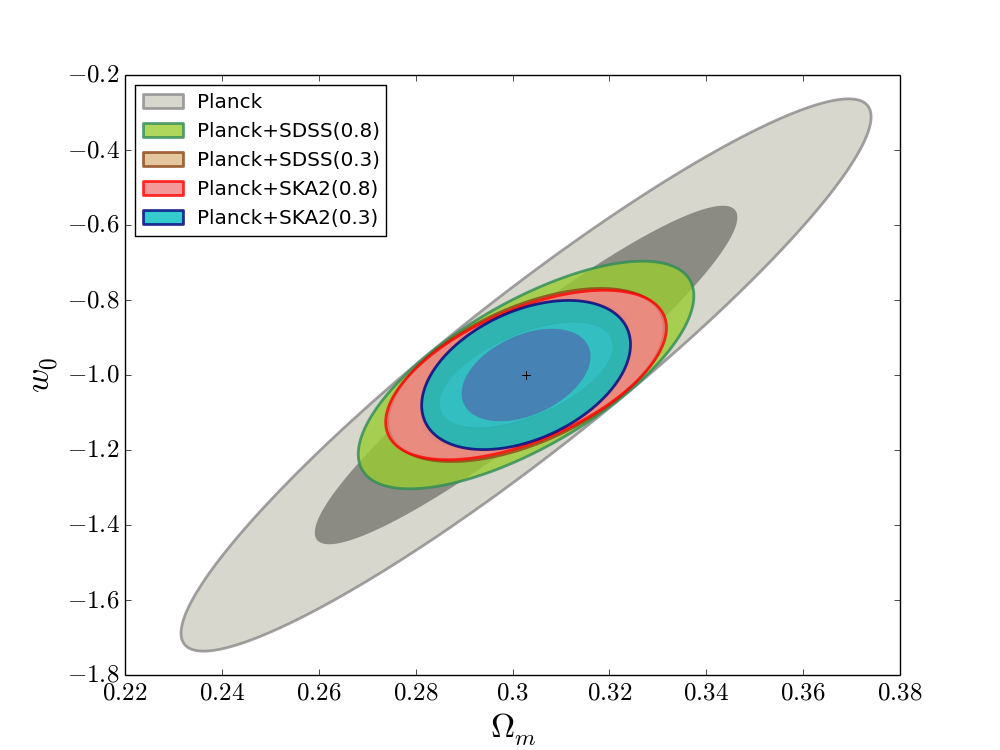}
\caption{\label{fig:ommw0all} Joint constraints on $\Omega_m-w_0$, marginalised over the other parameters, using Planck alone, Planck combined with SDSS, and Planck combined with SKA Phase 2. We use the dipole at separation 12\,Mpc/$h \leq d \leq$ 180\,Mpc/$h$. The numbers 0.3 and 0.8 refer to the value used for $\sigma_\kappa$. In the {\it left panel}, we consider the bias as a free parameter that is marginalised over. For SDSS we include three parameters, $b_1$, $b_2$ and $b_3$ (one for each sample), whereas for the SKA we have two parameters, $c_4$ and $c_5$ defined in Eq.~\eqref{evol_bias}. In the {\it right panel} we assume that the bias is known and fixed to its fiducial value.}
\end{figure*}

\begin{figure*}
\centering
\includegraphics[width=0.48\textwidth]{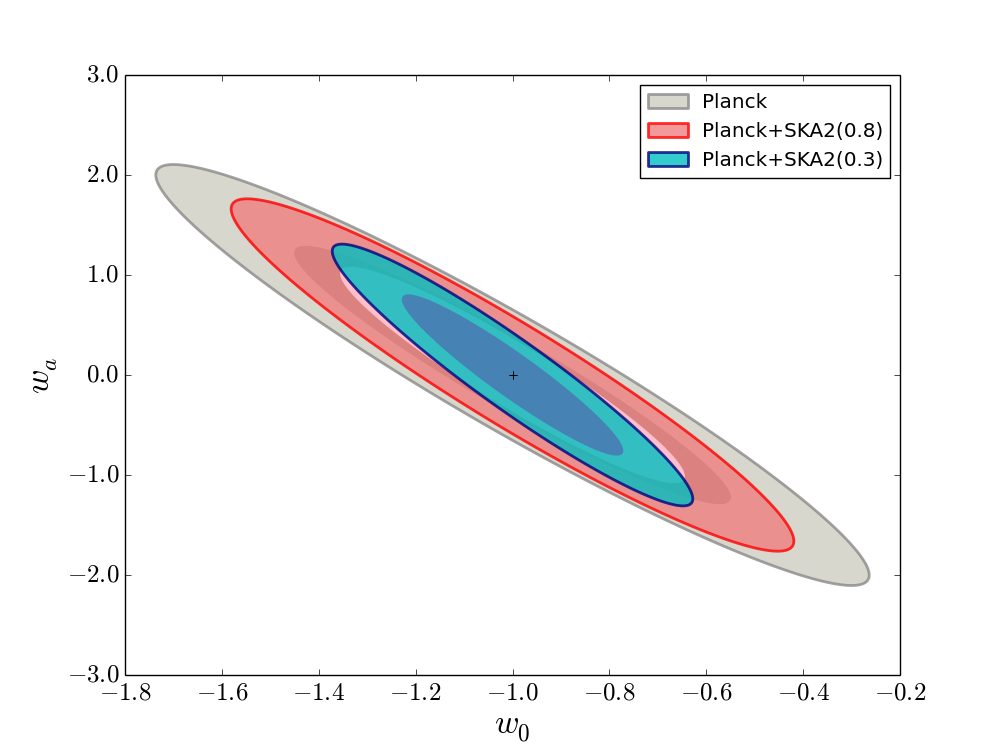}\hspace{0.5cm}
\includegraphics[width=0.48\textwidth]{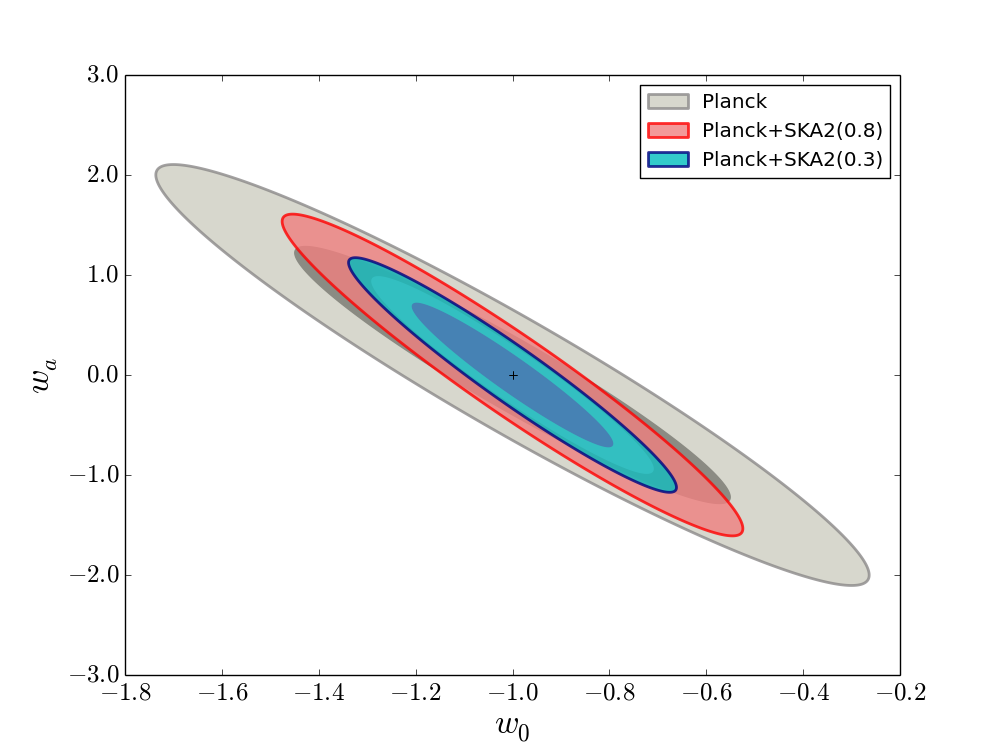}
\caption{\label{fig:waw0} Joint constraints on $w_0-w_a$, marginalised over the other parameters, using Planck alone and Planck combined with SKA Phase 2. We use the dipole at separation 12\,Mpc/$h \leq d \leq$ 180\,Mpc/$h$. The numbers 0.3 and 0.8 refer to the value used for $\sigma_\kappa$. In the {\it left panel}, the bias is described by two free parameters, $c_4$ and $c_5$, defined in Eq.~\eqref{evol_bias}, that are marginalised over. In the {\it right panel} we assume that the bias is known and we fix $c_4$ and $c_5$ to their fiducial values.}
\end{figure*}

\begin{figure*}
\centering
\includegraphics[width=0.48\textwidth]{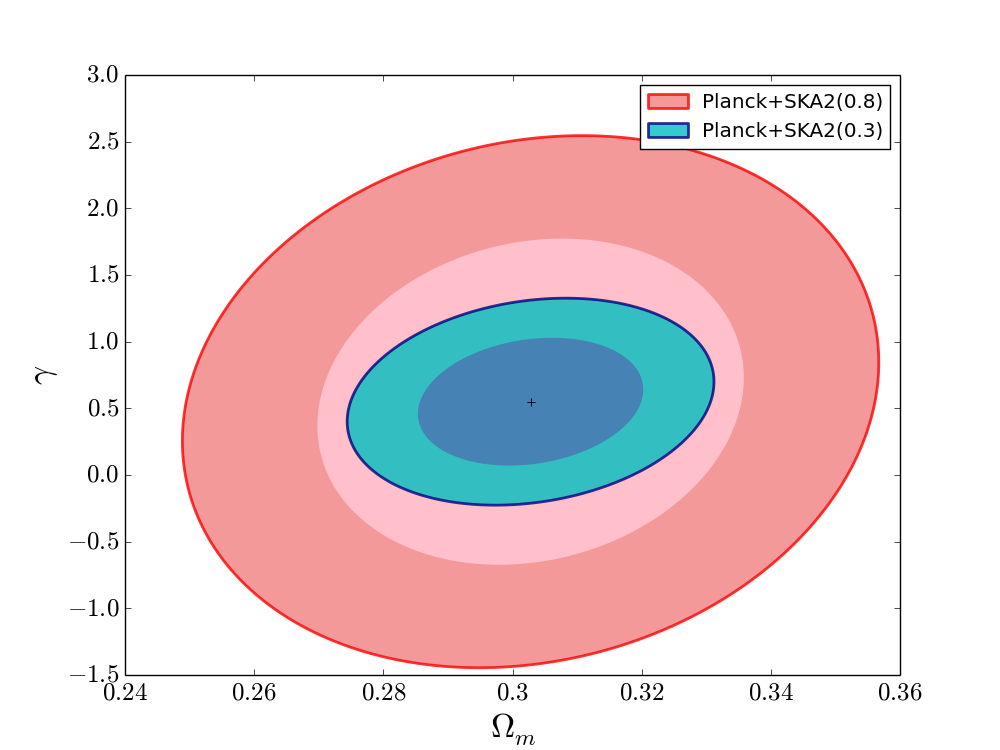}\hspace{0.5cm}
\includegraphics[width=0.48\textwidth]{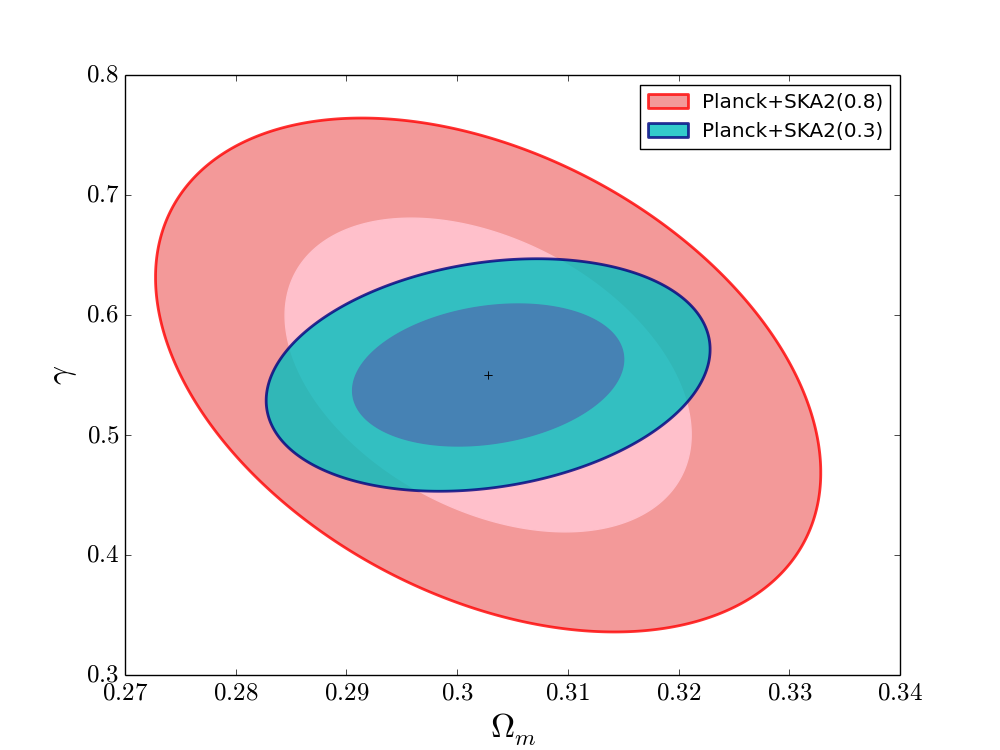}
\caption{\label{fig:ommgamma} Joint constraints on $\Omega_m-\gamma$, marginalised over the other parameters, using Planck combined with SKA Phase 2. We use the dipole at separation 12\,Mpc/$h \leq d \leq$ 180\,Mpc/$h$. The numbers 0.3 and 0.8 refer to the value used for $\sigma_\kappa$. In the {\it left panel}, the bias is described by two free parameters $c_4$ and $c_5$ defined in Eq.~\eqref{evol_bias}, that are marginalised over. In the {\it right panel} we assume that the bias is known and we fix $c_4$ and $c_5$ to their fiducial values.}
\end{figure*}

Since the Doppler magnification dipole should be detectable with both current and future experiments, we now forecast the constraints on cosmological parameters obtained from this measurement. We use scales between 12\,Mpc/$h$ and 180\,Mpc/$h$. The signal-to-noise of the dipole decreases relatively quickly with separation, and scales above 180\,Mpc/$h$ do not improve the constraints by much. At lower separation, on the other hand, the signal-to-noise increases significantly. We have however decided to remove scales below 12\,Mpc/$h$ since they are significantly affected by non-linearities~\footnote{The cut-off at 12\,Mpc/$h$ has been chosen by comparing the linear prediction for the dipole with the following proxy for the non-linear dipole: we have used linear Einstein's equations to relate the velocity to the density and modelled the non-linear density with the Halofit power spectrum. Above 12\,Mpc/$h$ the non-linear corrections obtained in this way are smaller than 5\%.}. To correctly model these non-linear scales, we should account not only for non-linearities in the density (which can be modelled using the non-linear Halofit power spectrum), but also for non-linearities in the velocity. This is beyond the scope of this paper.

We first consider constraints on five cosmological parameters: $\Omega_m, \Omega_b, h$, the primordial amplitude $A$ and the dark energy equation of state $w_0$ (assumed constant in redshift). For SDSS, we add three free bias parameters with fiducial values $b_1=1.17$ in the main sample of SDSS~\citep{Percival:2006gt, Cresswell:2008aa}, $b_2=1.77$ in LOWz and $b_3=1.89$ in CMASS~\citep{Gaztanaga:2015jrs}. For the SKA we can reasonably assume that the bias evolves smoothly over the five redshift bins. We model its evolution using 
\be\label{evol_bias}
b(z)=c_4\exp(c_5 z)\, ,
\ee
where $c_4$ and $c_5$ are two free parameters~\citep[see][]{Bull:2015lja}. 

In the left panel of Figure~\ref{fig:ommw0all} we show the joint constraints on $\Omega_m-w_0$, marginalised over the other cosmological parameters and bias parameters. We compare the constraints from using Planck alone, Planck combined with SDSS, and Planck combined with the SKA.\footnote{To include the Planck constraints (including CMB lensing power spectra to break the geometric degeneracy), we produced an approximate Fisher matrix by calculating the (inverse) covariance of the relevant cosmological parameters from the Planck 2015 \texttt{base\_w\_plikHM\_TT\_lowTEB\_post\_lensing} MCMC chains~\citep{Ade:2015xua}.} For each case we show how the constraints change when the error on the convergence goes from $\sigma_\kappa=0.8$ to $\sigma_\kappa=0.3$. We see that for SDSS, assuming $\sigma_\kappa=0.3$, the dipole already improves the constraints from Planck by 20 percent on $\Omega_m$ and 7 percent on $w_0$. With the SKA the improvement is even more significant, showing that the dipole genuinely adds valuable information on the growth of structure. 

Comparing with current constraints from redshift-space distortions, we see that the SKA constraints on $\Omega_m$ are similar to current BOSS constraints, whereas the constraints on $w_0$ are weaker by a factor 2~\citep[see e.g. Figure 13 of][]{Grieb:2016uuo}. The reason is that redshift-space distortions measure the monopole and the quadrupole, which are sensitive to different combinations of the bias and the growth rate. Combining those measurements allows one to separately constrain $b\sigma_8$ and $f\sigma_8$. The dipole, on the other hand, is sensitive to the combination $(b+3f/5)f\sigma_8$, and does not on its own allow separate constraints on the bias and the growth rate (see Eq.~\eqref{meandip}). In the right panel of Figure~\ref{fig:ommw0all}, we show how the constraints on $\Omega_m-w_0$ improve if we assume that the bias is perfectly known. We see that in this case the constraints from the dipole become much tighter for both SDSS and the SKA. The bias can be measured separately from higher-order correlation functions or lensing cross-correlations, making some of this increase in precision achievable in practice. Since they constrain different combinations of $f$ and $b$, redshift-space distortions and Doppler magnification dipole measurements could also be combined to break degeneracies between these parameters.

In Appendix~\ref{app:forecasts} we show the constraints on the other cosmological parameters from SDSS (Figure~\ref{fig:fisheropt}) and from the SKA (Figure~\ref{fig:fisherska}), marginalised over the bias parameters. We see that if the error on the convergence is as large as $\sigma_\kappa=0.8$ the dipole in SDSS adds almost nothing to the constraints from Planck. For $\sigma_\kappa=0.3$, the improvement over Planck alone is however non-negligible. Using the SKA, we see a significant improvement over Planck alone, for both values of $\sigma_\kappa$.

We then explore models beyond $\Lambda$CDM. First we let the equation of state evolve with time~\citep{Chevallier:2000qy, Linder:2002et} 
\be
w(a)=w_0+w_a(1-a)\, .
\ee 
In Figure~\ref{fig:waw0} we show the constraints on $w_0-w_a$ (marginalised over the other parameters) from Planck alone and Planck combined with the SKA. In the left panel we marginalise over the bias parameters, whereas in the right panel we fix the bias to its fiducial value. Comparing with the constraints from redshift-space distortions~\citep[see e.g. Figure 10 of][]{Grieb:2016uuo} we see that the Doppler magnification dipole provides slightly stronger constraints. We find that fixing the bias to its fiducial value improves the constraints by 20 percent on both $w_0$ and $w_a$. Note that the constraints on $w_0-w_a$ from the Doppler magnification dipole are similar to those obtained from shear measurements with the SKA Phase 2~\citep[see e.g. Figure 4 of][]{Harrison:2016stv}.

We then explore deviations from General Relativity by letting the growth rate evolve according to $f=(\Omega_m(a))^\gamma$, where $\gamma$ is a free parameter~\citep[in General Relativity $\gamma\simeq 0.55$, see e.g.][]{Wang:1998gt, Linder:2005in, Ferreira:2010sz}. Even though this parametrisation does not provide a description of all models beyond General Relativity, it is useful because it gives an easy way of assessing the potential of our observable to constrain modified gravity scenarios. In a forthcoming work we will study in detail how generic models of modified gravity affect the Doppler magnification dipole, and what kind of deviations from General Relativity can be constrained by this observable. In Figure~\ref{fig:ommgamma} we show the constraints on $\Omega_m-\gamma$ (marginalised over the other parameters) from Planck combined with the SKA.\footnote{Note that Planck does not provide constraints on $\gamma$, but does help to improve precision through the constraint on $\Omega_m$.} In the left panel we marginalise over the bias parameters, while in the right panel we fix the bias to its fiducial value.

When the bias is free, the constraints on $\gamma$ are weaker than those obtained from redshift-space distortions, see e.g. Figure 15 of~\citet{Grieb:2016uuo} (note however that the constraints are not directly comparable, since Figure 15 shows the constraints on $w_0-\gamma$). This reflects the fact that the dipole on its own does not allow us to constrain the bias and the growth rate separately: a change in the parameter $\gamma$ can therefore be reabsorbed into a change in the bias. Fixing the bias to its fiducial value breaks this degeneracy and consequently improves the constraints on $\gamma$ by a factor 7. This shows that combining measurement of the Doppler magnification dipole with bias measurements (for example from the monopole and quadrupole of redshift-space distortions) can potentially place stringent constraints on the growth rate.

Finally, we have explored how adding the octupole modifies the constraints on cosmological parameters. The octupole is potentially very interesting, as it does not depend on the galaxy bias; see Eq.~\eqref{meanoct}. However we find that adding the octupole leaves the constraints almost unchanged. This is because the signal-to-noise of the octupole is significantly lower than the one of the dipole, as shown in Figure~\ref{fig:SN_SKA}.

One could argue that the information contained in the Doppler magnification dipole is the same as the one in the monopole and quadrupole of redshift-space distortions, as they all probe the growth rate $f$. However, in addition to providing an independent measurement of the growth rate, the dipole also has the advantage of having a different dependence on scale. From Eq.~\eqref{nuell} we see that the shape of the dipole is determined by the integral of the power spectrum multiplied by $(\HH_0/k)j_1(kd)$. The monopole and quadrupole on the other hand contain an integral of the power spectrum multiplied by $j_0(kd)$ and $j_2(kd)$ respectively, without the $(\HH_0/k)$ suppression, see e.g.~\citet{Bonvin:2013ogt}. 
If the growth rate is independent of scale, as predicted by General Relativity, then the different scalings of the integrals is irrelevant, since the growth rate factors out. If the growth is scale-dependent, however, it will induce different signatures in the dipole than in the monopole and the quadrupole, due to the $\HH_0/k$ suppression. Combining the dipole with the monopole and quadrupole therefore provides a way of testing the consistency of a scale-independent growth rate.

\section{Comparison with other velocity estimators}
\label{sec:comp}

Measurements of peculiar velocities from galaxy surveys have a long history. In Section~\ref{sec:fisher} we compared the Doppler magnification dipole with velocity measurements from redshift-space distortions. Here we briefly discuss how our estimator compares with measurements of the velocity that combine redshift and distance~\citep{1977A&A....54..661T, Dressler:1987ny, Djorgovski:1987vx, Tonry:1999vt, Turnbull:2011ty, Tully:2013wqa, Springob:2014qja}, 
as well as with more recent propositions of measuring velocities using cross-correlations of galaxy populations with different biases~
\citep{Bonvin:2013ogt, Bonvin:2015kuc, Gaztanaga:2015jrs, Hall:2016bmm}. 

\subsection{Comparison with distance measurements}
\label{sec:distance}

In addition to measurements from redshift-space distortions, peculiar velocities have been measured through their effect on the distance to galaxies. More precisely, at low redshift we can write
\be
\label{distance}
\bV\cdot\bn=c z-H_0 r\, .
\ee
Combining redshift measurements with independent measurements of the distance $r$ therefore allows the galaxy's radial peculiar velocity to be measured directly. Various methods have been developed over the years to measure the distance to galaxies. For example, the Tully-Fisher relation~\citep{1977A&A....54..661T} allows us to measure distances to spiral galaxies, the $D_n-\sigma$ relation ($D_n$ being the luminous diameter and $\sigma$ the velocity dispersion) associated with the fundamental plane for elliptical galaxies provides a distance indicator for elliptical galaxies~\citep{Dressler:1987ny, Djorgovski:1987vx}, fluctuations of the surface brightness can be used to measure distances to early-type galaxies~\citep{Tonry:1999vt}, and flux measurements of supernovae allow us to measure their luminosity distance~\citep{Turnbull:2011ty}. Using Eq.~\eqref{distance} these distances can then be used to infer the peculiar velocities~\citep[see][for recent velocity catalogues]{Tully:2013wqa, Springob:2014qja}. These measurements are usually limited to low redshift. The first reason is that even a relatively small error on the distance generates a large error on the Hubble flow subtraction as the distance increases. For example, a 10\% error at 50\,Mpc$/h$ generates an error $H_0\delta r=500$\,km/s, i.e. of the order of magnitude of the peculiar velocity we want to measure. The second limitation comes from the fact that, as redshift increases, the contribution from gravitational lensing to the distance becomes more and more important~\citep[see e.g.][]{Bonvin:2005ps}, contaminating the measurement of peculiar velocities.

Our estimator is similar in essence to the methodology of Eq.~\eqref{distance}: we look at fluctuations in the size of galaxies (which are directly related to their distance) to infer the peculiar velocity. However by looking at cross-correlations between sizes and galaxy number counts, and by fitting for a dipole, we overcome the two problems associated with distance measurements. First, we get rid of the background part by averaging the sizes at fixed redshift and removing this average from the convergence. Second, as shown in Section~\ref{sec:estimator}, by fitting for a dipole we can efficiently remove the lensing contamination up to $z\simeq 0.5$, and even at high redshift $z\simeq 1$ we can reduce the impact of gravitational lensing drastically. These improvements do not directly measure the radial velocity as in Eq.~\eqref{distance} however, but rather its correlation with density fluctuations.  

\subsection{Comparison with the dipole of $\langle \Delta \Delta\rangle$}  

Another method to measure peculiar velocities has been proposed recently in~\citet{Bonvin:2013ogt, Bonvin:2015kuc, Gaztanaga:2015jrs}. The idea is to cross-correlate different populations of galaxies with different biases and to fit for a dipole in the cross-correlation. This allows us to isolate the following combination of velocities and the gradient of the potential in the number counts:
\begin{align}
\label{deltarel}
\Delta^{\rm dip}=&\left[1-\frac{\dot\HH}{\HH^2}-\frac{2}{r\HH}+5s\left(1-\frac{1}{r\HH} \right)\right]\bV\cdot\bn\\
&+\frac{1}{\HH}\partial_r\Psi+ \frac{1}{\HH}\dot{\bV}\cdot\bn\, .\nonumber
\end{align}
This dipole has a lower signal-to-noise than the dipole of $\langle\Delta \kappa\rangle$~\citep[see][]{Bonvin:2015kuc}, and will be challenging to measure in current galaxy surveys -- the cumulative signal-to-noise in the main sample of SDSS galaxies is 2.4. It should be robustly detected in future galaxy surveys though, such as DESI (where the signal-to-noise is 7.4). With SKA Phase 2, we should be able to detect it with a signal-to-noise of $\sim$100~\citep{Hall:2016bmm}. From Eq.~\eqref{deltarel}, we see that the dipole of $\langle\Delta \Delta\rangle$ measures a different combination of velocities than the dipole of $\langle\Delta \kappa\rangle$, see Eq.~\eqref{kappav}. Furthermore, the dipole of $\langle\Delta \Delta\rangle$ is also sensitive to the gradient of $\Psi$. Combining the two dipoles would therefore allow us to test the validity of the Euler equation in a model-independent way, i.e. to test if galaxies move according to the gravitational potential $\Psi$ or if they are affected by an additional force. 

\section{Conclusion}
\label{sec:conclusion}

Peculiar velocities are useful for testing the consistency of General Relativity, as they allow us to directly measure the rate at which structures grow with time. Combined with density measurements, velocities therefore provide useful constraints on cosmological models beyond $\Lambda$CDM. Various methods have been proposed over the years to measure peculiar velocities from large-scale structure observations. A key approach consists in looking at how peculiar velocities change the apparent distance between us and nearby objects~\citep{1977A&A....54..661T, Dressler:1987ny, Djorgovski:1987vx, Tonry:1999vt, Turnbull:2011ty, Tully:2013wqa, Springob:2014qja}. By combining distance measurements with redshift information, one can measure the radial component of peculiar velocities. This method has delivered useful measurements of galaxies' velocities, but has the disadvantage of being restricted to low redshifts, where the scatter in the distance measurements does not wash out the signal.

Another fruitful method, which has been used extensively during the last decades, consists in looking at how peculiar velocities change the amplitude of the two-point correlation function of galaxies (or of its Fourier transform, the power spectrum), via the so-called redshift-space distortions~~\citep{Kaiser:1987qv, 1989MNRAS.236..851L, 1992ApJ...385L...5H}. The origin of the distortions is the same as before: velocities change the apparent distance to the galaxies, and since we use distances to pixelise our sky, they change the size of the redshift bins in which we count how many galaxies we have. As a consequence, the number of galaxies that we detect per pixel is modified by peculiar velocities. A whole machinery has been developed over the years to extract velocity measurements from the two-point correlation function and the power spectrum, giving rise to valuable constraints on cosmological parameters~\citep[see e.g.][]{Hawkins:2002sg, Zehavi:2004zn, Guzzo:2008ac, Cabre:2008sz, Song:2010kq, Samushia:2013yga, Chuang:2013wga, Satpathy:2016tct, Beutler:2016arn}.

Here, we have proposed an alternative method: since velocities change the observed distance to galaxies, they also change their apparent size. Consequently, measurements of the convergence field are automatically affected by peculiar velocities~\citep{Bonvin:2008ni, Bolejko:2012uj, Bacon:2014uja}. In this paper we constructed an estimator to measure the velocities using this effect. We have shown that by correlating the convergence with the number counts of galaxies, and by fitting for a dipole, we can isolate the velocity contribution from the gravitational lensing contribution up to a redshift of $\sim 0.5$. This method therefore provides a competitive alternative to other velocity probes. We have shown that the signal-to-noise of the dipole is sufficiently large to be detected in current optical surveys. We also forecasted the signal-to-noise for the future DESI and SKA2 HI galaxy surveys, showing that the dipole will be robustly detected in these samples. Finally, we computed the expected constraints on cosmological parameters for SDSS and the SKA, demonstrating the potential of this new observable to test the growth of structure.

The information from size measurements is expected to be similar to that from redshift-space distortions, as in both cases the effect is due to the impact of velocities on distances. Measuring sizes is very different from counting objects however, and so we expect the two observables to be affected differently by uncertainties. Moreover, since the Doppler magnification dipole has a different scale dependence than redshift-space distortions, it allows us to test the consistency of a scale-independent growth rate. The convergence dipole therefore provides a new and competitive method to measure peculiar velocities from large-scale structure surveys.

\section*{Acknowledgements}
It is a pleasure to thank Elisa Chisari for interesting discussions. CB acknowledges support by the Swiss National Science Foundation. DB and RM are supported by the STFC (UK, Grant ST/K00090X/1). SA and RM are supported by the South African SKA Project. RM is also supported by the NRF (South Africa). PB's research was supported by an appointment to the NASA Postdoctoral Program at the Jet Propulsion Laboratory, California Institute of Technology, administered by Universities Space Research Association under contract with NASA.

\appendix

\section{Mean of the dipole and octupole estimators}
\label{app:mean}

We calculate the mean of the dipole and octupole estimators~\eqref{estimatordip} and~\eqref{estimatoroct}. In the continuous limit the sum over pixels becomes
\be
\sum_i=\frac{1}{\ell^3_p}\int d^3\bx_i\quad\mbox{and}\quad \delta_K(d_{ij}-d)=\ell_p\delta_D(d_{ij}-d)\, ,
\ee
where $\ell_p$ is the size of the cubic pixels in which we measure the convergence and the number counts. The mean of the dipole then becomes
\be
\langle \hat \xi_{\rm dip}\rangle(d)= \frac{\norm}{\ell_p^5}\int d^3\bx_i \int d^3\bx_j \langle\Delta_i \kappa_j\rangle \cos\beta_{ij}\delta_D(d_{ij}-d)\, .
\ee
Since the Universe is statistically homogeneous and isotropic, we can fix the position of the pixel $i$ and then multiply by the volume of the survey $V$ to account for the integral over $\bx_i$. The integral over $\bx_j$ can be expressed in spherical coordinates. By isotropy, the two-point function does not depend on the azimuthal angle $\varphi$, so the mean is
\be
\langle\hat \xi_{\rm dip}\rangle(d)= \frac{2\pi\norm V d^2}{\ell_p^5}\int_0^\pi d\beta \sin\beta\cos\beta \langle\Delta \kappa\rangle \, .
\ee
Inserting the expression for $\langle\Delta \kappa\rangle$ and integrating over $\beta$ we find
\begin{align}
\langle\hat{\xi}_{\rm dip}\rangle(d)=&\frac{4\pi\norm V d^2}{3\ell_p^5}\frac{\HH(z)}{\HH_0}f(z)\\
&\times \left(1-\frac{1}{\HH(z)r} \right)\left(b(z)+\frac{3f(z)}{5} \right)\nu_1(d)\, .\nonumber
\end{align}
Choosing the normalisation
\be
\label{normapp}
\norm=\frac{3}{4\pi}\frac{\ell_p^5}{d^2V}\, ,
\ee
we obtain Eq.~\eqref{meandip}. A similar calculation for the octupole gives Eq.~\eqref{meanoct}.

\section{Variance}
\label{app:variance}

To calculate the second term in Eq.~\eqref{variance} we follow the derivation presented in~\citet{Hall:2016bmm}. In the continuous limit, we have
\begin{align}
\label{var2app}
{\rm var}_2= &\frac{9\ell_p^3\sigma_\kappa^2}{16\pi^2 V^2 d^2 d'^2}\int d^3\bx_i\int d^3\bx_j\int d^3\bx_aC^\Delta_{ia}\\
&\times \cos\beta_{ij}\cos\beta_{aj}\delta_D(d_{ij}-d)\delta_D(d_{aj}-d')\, .\nonumber
\end{align}
We make a change of variables $\by_i=\bx_i-\bx_j$ and $\by_a=\bx_a-\bx_j$. The integral over $\bx_j$ becomes trivial, and gives the volume of the survey $V$. In the distant observer approximation, where we have one fixed line-of-sight $\bn$, Eq.~\eqref{var2app} becomes
\begin{align}
\label{vary}
{\rm var}_2=& \frac{9\ell_p^3\sigma_\kappa^2}{4(2\pi)^5 V d^2 d'^2}\int d^3\bk \ P(k,z)
\Bigg[b^2+\frac{2bf}{3}+\frac{f^2}{5}\\
& +\left(\frac{4bf}{3}+\frac{4f^2}{7} \right)P_2(\hat\bk\cdot\bn)+\frac{8f^2}{35} P_4(\hat\bk\cdot\bn)\Bigg]\nonumber\\
&\times\int d^3\by_i e^{-i\bk\cdot\by_i}\cos\beta_{\by_i}\delta_D(y_i-d)\nonumber\\
&\times\int d^3\by_a e^{i\bk\cdot\by_a}\cos\beta_{\by_a}\delta_D(y_a-d')\, .\nonumber
\end{align}
The integrals over $\by_i$ and $\by_a$ can be performed analytically, giving rise to
\be
16\pi^2 d^2 d'^2 \cos(\bn\cdot\bk)^2 j_1(kd)j_1(kd')\, .
\ee
Inserting this into~\eqref{vary} we obtain
\begin{align}
&{\rm var}_2=\frac{9}{4\pi^2}\frac{\ell_p^3}{V}\sigma^2_\kappa\int \!dk k^2 P(k,z)j_1(dk)j_1(kd')\int_{-1}^1\!d\mu \mu^2\\
&\times\left[b^2+\frac{2bf}{3}+\frac{f^2}{5} +\left(\frac{4bf}{3}+\frac{4f^2}{7} \right)P_2(\mu)+\frac{8f^2}{35} P_4(\mu)\right]\, .\nonumber
\end{align}
Performing the integral over $\mu$, we obtain Eq.~\eqref{var2}.

\section{Variance for galaxies with different $\sigma_\kappa$}
\label{app:kappamixed}

In Section~\ref{sec:variance} and Appendix~\ref{app:variance} the variance was calculated assuming that the size of all galaxies can be measured with the same error $\sigma_\kappa$. In reality, surveys are composed of both early-type (elliptical) galaxies, for which $\sigma_\kappa\sim 0.3$, and late-type (spiral) galaxies for which $\sigma_\kappa\sim 0.8$, as discussed in~\cite{Alsing:2014fya}. To calculate the variance in this case, we assume that in each pixel $j$ we measure either only the convergence from elliptical galaxies, with an error $\sigma_\kappa^{\rm E}=0.3$, or only from spiral galaxies, with an error $\sigma_\kappa^{\rm S}=0.8$. The estimator for the dipole becomes
\be
\label{estimatormix}
\hat\xi_{\rm dip}(d)=\norm \sum_{i}\left(\sum_{j\in {\rm E}} \Delta_i \kappa^{\rm E}_j+\sum_{j\in {\rm S}} \Delta_i \kappa^{\rm S}_j\right) \cos\beta_{ij}\delta_K(d_{ij}-d)\, ,
\ee
where $\kappa_j^{\rm E}=\kappa_j+\sigma_\kappa^{\rm E}$ and $\kappa_j^{\rm S}=\kappa_j+\sigma_\kappa^{\rm S}$, and $\kappa_j$ is the true convergence in pixel $j$. The noise cancels on average, so that
\be
\langle\Delta_i\kappa^{E}_j \rangle=\langle\Delta_j\kappa^{S}_j \rangle=\langle\Delta_j\kappa_j \rangle\, ,
\ee
and we recover the mean of the estimator given by Eq.~\eqref{meandip}.

The variance of the estimator is different, however. Using Eq.~\eqref{estimatormix} we obtain
\begin{align}
&{\rm var}\Big(\hat \xi_{\rm dip}\Big)=\left\langle\left(\hat \xi_{\rm dip}\right)^2\right\rangle-\left\langle\hat \xi_{\rm dip}\right\rangle^2=\frac{9\ell_p^{10}}{16\pi^2 V^2 d^2 d'^2}\nonumber\\
&\times\sum_{ia} \sum_{AB\in\{{\rm E, S}\}} \sum_{j\in A}\sum_{b\in B}\Big[ \langle \Delta_i \kappa^{ A}_j \Delta_a \kappa^{ B}_b\rangle-\langle \Delta_i \kappa^{ A}_j\rangle \langle\Delta_a \kappa^{ B}_b\rangle \Big]\nonumber\\
&\quad\times \cos\beta_{ij}\cos\beta_{ab}\delta_K(d_{ij}-d)\delta_K(d_{ab}-d')\, . \label{vargenmix}
\end{align}
As explained in Section~\ref{sec:variance}, the dominant contributions to the variance are from the auto-correlation of $\Delta$ and $\kappa$, so that Eq.~\eqref{vargenmix} becomes
\begin{align}
&{\rm var}\Big(\hat \xi_{\rm dip}\Big)=\frac{9\ell_p^{10}}{16\pi^2 V^2 d^2 d'^2}
\sum_{ia} \sum_{AB\in\{{\rm E, S}\}} \sum_{j\in A}\sum_{b\in B}\nonumber\\
&\times\langle \Delta_i \Delta_a\rangle\langle\kappa^{ A}_j  \kappa^{ B}_b\rangle\times \cos\beta_{ij}\cos\beta_{ab}\delta_K(d_{ij}-d)\delta_K(d_{ab}-d')\, . \label{varmix}
\end{align}
Since we have assumed that in one pixel we have either elliptical or spiral galaxies (but not both), we can write
\be
\label{kappaAB}
\langle\kappa^{ A}_j  \kappa^{ B}_b\rangle=\big(\sigma_\kappa^A\big)^2\delta_{AB}\delta_{jb}\, ,
\ee
where, as explained in Section~\ref{sec:variance}, we can neglect the cosmic variance contribution $C^\kappa_{ij}$. Inserting~\eqref{kappaAB} and~\eqref{DeltaDelta} into~\eqref{varmix} we obtain
\begin{align}
{\rm var}\Big(\hat \xi_{\rm dip}\Big)&=\frac{9\ell_p^{10}}{16\pi^2 V^2 d^2 d'^2}\sum_{A\in\{{\rm E, S}\}} \sum_{j\in A}
\big(\sigma_\kappa^A\big)^2\label{varmixdiscrete}\\
&\times\Bigg[ \frac{1}{\delta \bar n}\sum_{i} \cos^2\beta_{ij}
\delta_K(d_{ij}-d)\delta_K(d-d')\nonumber\\
& +\sum_{ia}C^\Delta_{ia}\cos\beta_{ij}\cos\beta_{aj}\delta_K(d_{ij}-d)\delta_K(d_{aj}-d')\Bigg]\, .\nonumber
\end{align}
Expression~\eqref{varmixdiscrete} can be calculated in the continuous limit following the same steps as described in Section~\ref{sec:variance} and Appendix~\ref{app:variance}. The only difference is that the sum over pixels $j$ runs separately over elliptical and spiral galaxies. We can rewrite it as 
\be
\sum_{j\in A}\rightarrow \frac{1}{\ell_p^3}\int d^3\bx_{j\in A}= \frac{1}{\ell_p^3} V \frac{N_{\rm tot}^A}{N_{\rm tot}}\, ,
\ee
where $N_{\rm tot}^A$ denotes the number of galaxies of type $A$. With this, the two contributions to the variance take the simple form
\be
{\rm var}_1\Big(\hat \xi_{\rm dip}\Big)=\frac{3}{4\pi}\frac{\ell_p^2}{d^2}\frac{\big(\sigma^{\rm mixed}_\kappa\big)^2}{N_{\rm tot}}\delta_K(d-d')\, , \label{var1mix}
\ee
\begin{align}
{\rm var}_2\Big(\hat \xi_{\rm dip}\Big)=&\frac{9}{2\pi^2}\frac{\ell_p^3}{V}\big(\sigma^{\rm mixed}_\kappa\big)^2\left(\frac{b^2}{3}+\frac{2bf}{5}+\frac{f^2}{7} \right)\label{var2mix}\\
&\times\int dk k^2 P(k,z)j_1(kd)j_1(kd')\, , \nonumber
\end{align}
with 
\be
\big(\sigma^{\rm mixed}_\kappa\big)^2=\frac{N^{\rm E}_{\rm tot}}{N_{\rm tot}}\big(\sigma^{\rm E}_\kappa\big)^2
+\frac{N^{\rm S}_{\rm tot}}{N_{\rm tot}}\big(\sigma^{\rm S}_\kappa\big)^2\, .
\ee
In a sample with 50\% elliptical galaxies and 50\% spiral galaxies, we obtain $\sigma^{\rm mixed}_\kappa=0.6$. The signal-to-noise in this case is therefore reduced by a factor 2 with respect to the optimal scenario, where all galaxies are assumed to be measured with $\sigma_\kappa=0.3$.

\section{Average over thick shells}
\label{app:thick}

In Section~\ref{sec:variance}, the mean and variance of the dipole and octupole were calculated at fixed separations $d$ and $d'$ between pixels. 
In practice we may want to average the signal over a range of separations $d_{\rm min}\leq d \leq d_{\rm max}$. In this case, the estimator for the dipole reads
\be
\hat \xi_{\rm dip}(d)=\norm \sum_{ij} \Delta_i \kappa_j \cos\beta_{ij}\Theta(d_{ij}-\dmin)\Theta(\dmax-d_{ij})\, ,\label{estimatorthick}
\ee
where $\Theta$ denotes the Heaviside step function $\Theta(x)=1$ if $x\geq 0$ and zero elsewhere. In the continuous limit, the mean of Eq.~\eqref{estimatorthick} becomes
\begin{align}
\langle\hat{\xi}_{\rm dip}\rangle=&\frac{\HH(z)}{\HH_0}f(z)\left(1-\frac{1}{\HH(z)r} \right)\left(b(z)+\frac{3f(z)}{5} \right)\\
&\times \frac{1}{\ell_p \bar d^2}\int_{\dmin}^{\dmax} \!ds \, s^2 \nu_1(s)\, ,\nonumber
\end{align}
where $\bar d$ is the mean separation between $\dmin$ and $\dmax$.

The first contribution to the variance reads
\begin{align}
&{\rm var}_1=\frac{9\ell_p^{10}}{16\pi^2 V^2 \bar d^2 \bar d'^2} \frac{\sigma_\kappa^2}{\delta\bar n}\sum_{ij}(\cos\beta_{ij})^2\\
&\times \Theta(d_{ij}-\dmin)\Theta(\dmax-d_{ij})\Theta(d_{ij}-\dmin')\Theta(\dmax'-d_{ij})\, .\nonumber
\end{align}
Taking the continuous limit, we obtain
\be
{\rm var}_1=\frac{3}{4\pi}\frac{\ell_p(\tilde{d}_{\rm max}^3-\tilde{d}_{\rm min}^3)}{3N_{\rm tot}\bar d^2 \bar d'^2}\sigma^2_\kappa\, ,
\ee
where $\tilde{d}_{\rm min}={\rm max}\big(\dmin, \dmin'\big)$ and $\tilde{d}_{\rm max}={\rm min}\big(\dmax, \dmax'\big)$.
The second contribution to the variance reads
\begin{align}
&{\rm var}_2=\frac{9\ell_p^{10}}{16\pi^2 V^2 \bar d^2 \bar d'^2} \sigma^2_\kappa\sum_{ija}C^\Delta_{ia}\cos\beta_{ij}\cos\beta_{aj}\\
&\times \Theta(d_{ij}-\dmin)\Theta(\dmax-d_{ij})\Theta(d_{aj}-\dmin')\Theta(\dmax'-d_{aj})\, .\nonumber
\end{align}
Following the same steps as in Appendix~\ref{app:variance}, we obtain
\begin{align}
&{\rm var}_2\Big(\hat \xi_{\rm dip}\Big)=\frac{9\ell_p}{2\pi^2V\bar d^2\bar d'^2}\sigma^2_\kappa\left(\frac{b^2}{3}+\frac{2bf}{5}+\frac{f^2}{7} \right)\label{var2thick}\\
&\times\int dk k^2 P(k,z)\int_{\dmin}^{\dmax}ds\, s^2 j_1(ks)\int_{\dmin'}^{\dmax'}ds' s'^2j_1(ks')\, . \nonumber
\end{align}
Similar expressions can be derived for the octupole.

\begin{figure*}
\centering
\includegraphics[width=0.52\textwidth]{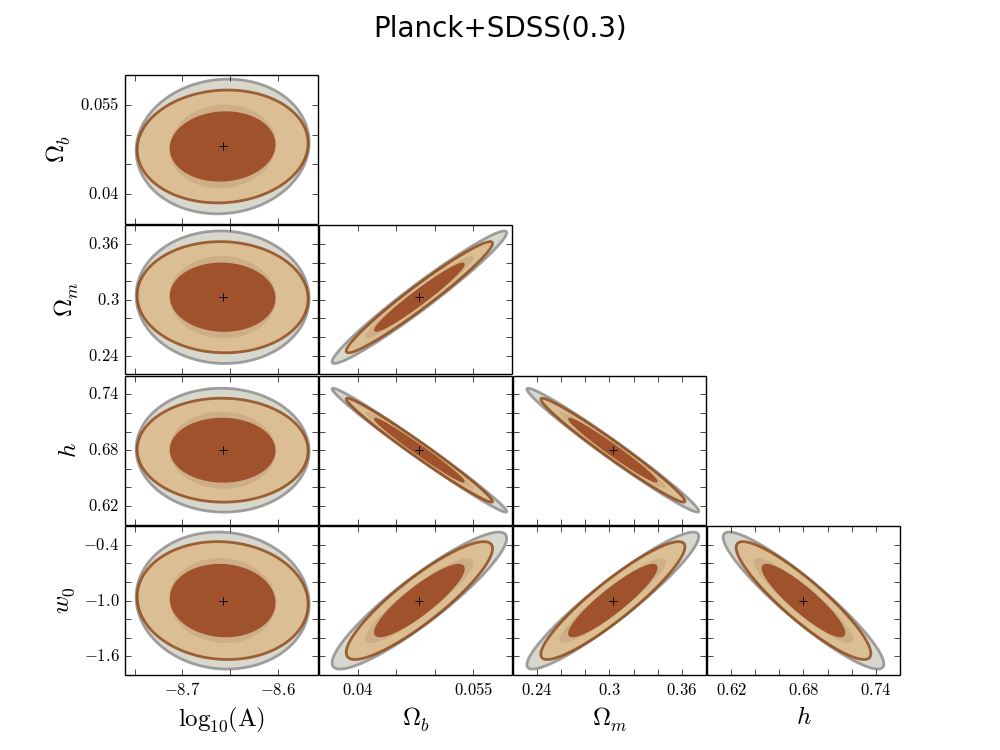}\hspace{-0.9cm}
\includegraphics[width=0.52\textwidth]{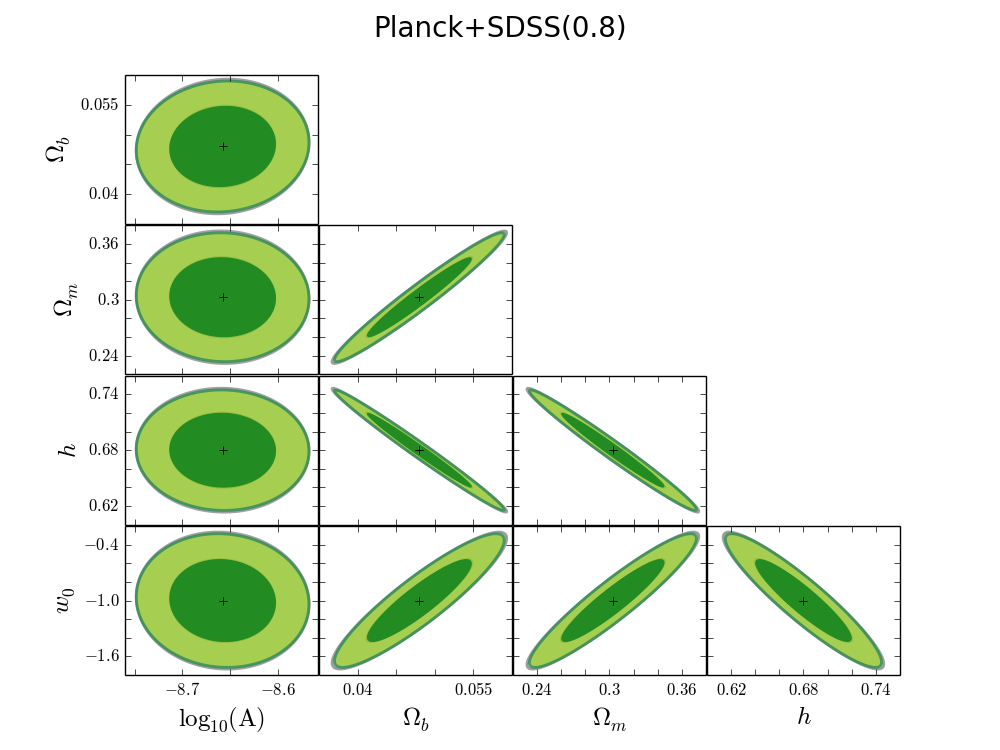}
\caption{\label{fig:fisheropt} Joint constraints on five cosmological parameters, using Planck alone (grey outside contours) and Planck combined with SDSS (colour inside contours). We use scales between 12\,Mpc/$h$ and 180\,Mpc/$h$ and marginalise over the bias parameters $b_1, b_2$ and $b_3$. We show constraints for $\sigma_\kappa=0.3$ (left panel) and $\sigma_\kappa=0.8$ (right panel).}
\end{figure*}

\begin{figure*}
\centering
\includegraphics[width=0.52\textwidth]{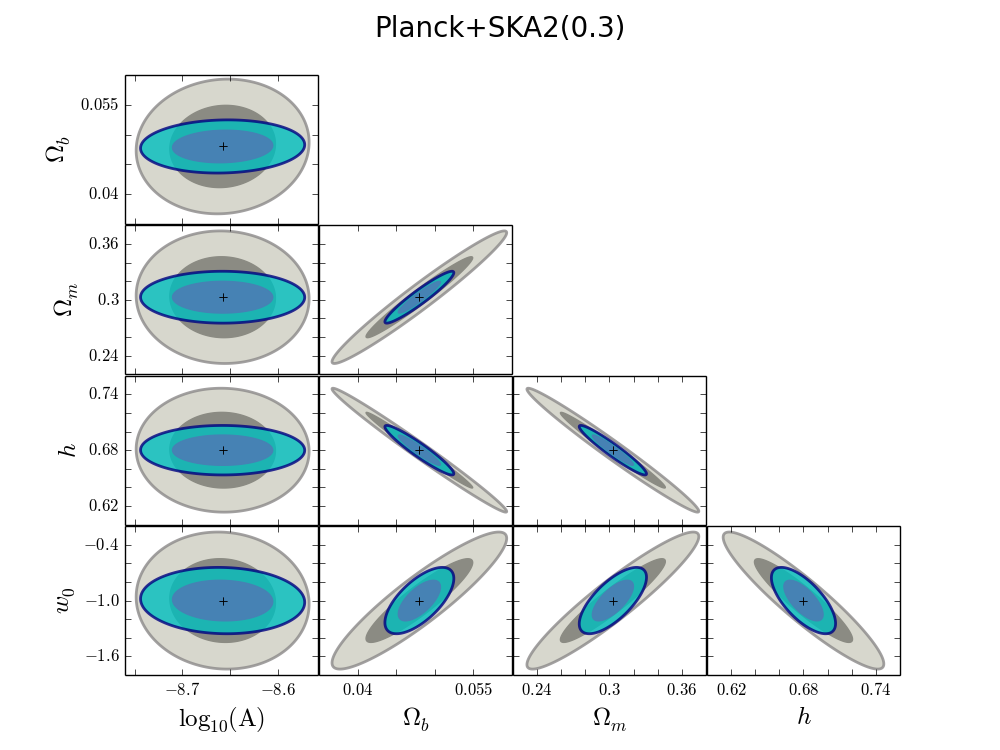}\hspace{-0.9cm}
\includegraphics[width=0.52\textwidth]{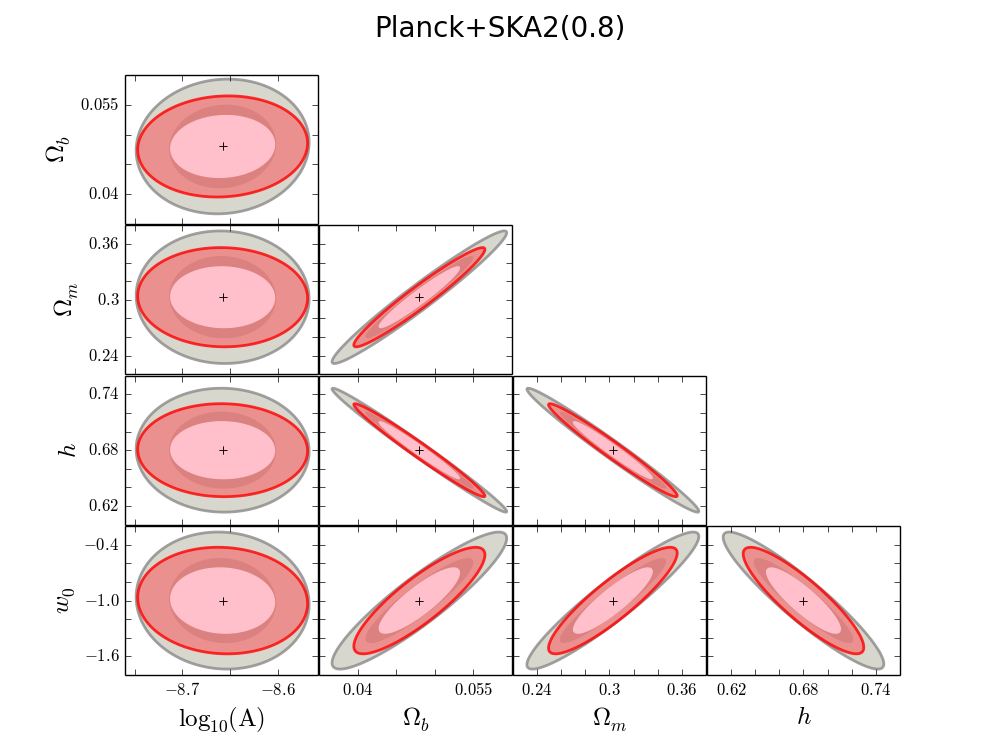}
\caption{\label{fig:fisherska} Joint constraints on five cosmological parameters, using Planck alone (grey outside contours) and Planck combined with SKA (colour inside contours). We use scales between 12\,Mpc/$h$ and 180\,Mpc/$h$ and marginalise over the bias parameters $c_4$ and $c_5$. We show constraints for $\sigma_\kappa=0.3$ (left panel) and $\sigma_\kappa=0.8$ (right panel).}
\end{figure*}

\section{Fisher forecasts}
\label{app:forecasts}

In Figures~\ref{fig:fisheropt} and~\ref{fig:fisherska} we show the constraints on $\Omega_m, \Omega_b, h, w_0$ and $A$ using SDSS and the SKA Phase 2.


\bibliographystyle{mnras}
\bibliography{bibdoppler.bib} 

\begin{thebibliography}{}
\makeatletter
\relax
\def\mn@urlcharsother{\let\do\@makeother \do\$\do\&\do\#\do\^\do\_\do\%\do\~}
\def\mn@doi{\begingroup\mn@urlcharsother \@ifnextchar [ {\mn@doi@}
  {\mn@doi@[]}}
\def\mn@doi@[#1]#2{\def\@tempa{#1}\ifx\@tempa\@empty \href
  {http://dx.doi.org/#2} {doi:#2}\else \href {http://dx.doi.org/#2} {#1}\fi
  \endgroup}
\def\mn@eprint#1#2{\mn@eprint@#1:#2::\@nil}
\def\mn@eprint@arXiv#1{\href {http://arxiv.org/abs/#1} {{\tt arXiv:#1}}}
\def\mn@eprint@dblp#1{\href {http://dblp.uni-trier.de/rec/bibtex/#1.xml}
  {dblp:#1}}
\def\mn@eprint@#1:#2:#3:#4\@nil{\def\@tempa {#1}\def\@tempb {#2}\def\@tempc
  {#3}\ifx \@tempc \@empty \let \@tempc \@tempb \let \@tempb \@tempa \fi \ifx
  \@tempb \@empty \def\@tempb {arXiv}\fi \@ifundefined
  {mn@eprint@\@tempb}{\@tempb:\@tempc}{\expandafter \expandafter \csname
  mn@eprint@\@tempb\endcsname \expandafter{\@tempc}}}

\bibitem[\protect\citeauthoryear{Ade et~al.}{Ade et~al.}{2015}]{Ade:2015xua}
Ade P.,  et~al., 2015

\bibitem[\protect\citeauthoryear{Alsing, Kirk, Heavens  \& Jaffe}{Alsing
  et~al.}{2015}]{Alsing:2014fya}
Alsing J.,  Kirk D.,  Heavens A.,   Jaffe A.,  2015, \mn@doi [Mon. Not. Roy.
  Astron. Soc.] {10.1093/mnras/stv1249}, 452, 1202

\bibitem[\protect\citeauthoryear{Anderson et~al.}{Anderson
  et~al.}{2014}]{Anderson:2013zyy}
Anderson L.,  et~al., 2014, \mn@doi [Mon. Not. Roy. Astron. Soc.]
  {10.1093/mnras/stu523}, 441, 24

\bibitem[\protect\citeauthoryear{Bacon, Andrianomena, Clarkson, Bolejko  \&
  Maartens}{Bacon et~al.}{2014}]{Bacon:2014uja}
Bacon D.~J.,  Andrianomena S.,  Clarkson C.,  Bolejko K.,   Maartens R.,  2014,
  \mn@doi [Mon. Not. Roy. Astron. Soc.] {10.1093/mnras/stu1270}, 443, 1900

\bibitem[\protect\citeauthoryear{Bernardeau, Bonvin  \& Vernizzi}{Bernardeau
  et~al.}{2010}]{Bernardeau:2009bm}
Bernardeau F.,  Bonvin C.,   Vernizzi F.,  2010, \mn@doi [Phys. Rev.]
  {10.1103/PhysRevD.81.083002}, D81, 083002

\bibitem[\protect\citeauthoryear{Bernardeau, Bonvin, Van~de Rijt  \&
  Vernizzi}{Bernardeau et~al.}{2012}]{Bernardeau:2011tc}
Bernardeau F.,  Bonvin C.,  Van~de Rijt N.,   Vernizzi F.,  2012, \mn@doi
  [Phys. Rev.] {10.1103/PhysRevD.86.023001}, D86, 023001

\bibitem[\protect\citeauthoryear{Beutler et~al.}{Beutler
  et~al.}{2016}]{Beutler:2016arn}
Beutler F.,  et~al., 2016, Submitted to: Mon. Not. Roy. Astron. Soc.

\bibitem[\protect\citeauthoryear{Bolejko, Clarkson, Maartens, Bacon, Meures  \&
  Beynon}{Bolejko et~al.}{2013}]{Bolejko:2012uj}
Bolejko K.,  Clarkson C.,  Maartens R.,  Bacon D.,  Meures N.,   Beynon E.,
  2013, \mn@doi [Phys. Rev. Lett.] {10.1103/PhysRevLett.110.021302}, 110,
  021302

\bibitem[\protect\citeauthoryear{Bonvin}{Bonvin}{2008}]{Bonvin:2008ni}
Bonvin C.,  2008, \mn@doi [Phys. Rev.] {10.1103/PhysRevD.78.123530}, D78,
  123530

\bibitem[\protect\citeauthoryear{Bonvin \& Durrer}{Bonvin \&
  Durrer}{2011}]{Bonvin:2011bg}
Bonvin C.,  Durrer R.,  2011, \mn@doi [Phys. Rev.]
  {10.1103/PhysRevD.84.063505}, D84, 063505

\bibitem[\protect\citeauthoryear{Bonvin, Durrer  \& Gasparini}{Bonvin
  et~al.}{2006}]{Bonvin:2005ps}
Bonvin C.,  Durrer R.,   Gasparini M.~A.,  2006, \mn@doi [Phys. Rev.]
  {10.1103/PhysRevD.85.029901, 10.1103/PhysRevD.73.023523}, D73, 023523

\bibitem[\protect\citeauthoryear{Bonvin, Hui  \& Gaztanaga}{Bonvin
  et~al.}{2014}]{Bonvin:2013ogt}
Bonvin C.,  Hui L.,   Gaztanaga E.,  2014, \mn@doi [Phys. Rev.]
  {10.1103/PhysRevD.89.083535}, D89, 083535

\bibitem[\protect\citeauthoryear{Bonvin, Hui  \& Gaztanaga}{Bonvin
  et~al.}{2016}]{Bonvin:2015kuc}
Bonvin C.,  Hui L.,   Gaztanaga E.,  2016, \mn@doi [JCAP]
  {10.1088/1475-7516/2016/08/021}, 1608, 021

\bibitem[\protect\citeauthoryear{Bull}{Bull}{2016}]{Bull:2015lja}
Bull P.,  2016, \mn@doi [Astrophys. J.] {10.3847/0004-637X/817/1/26}, 817, 26

\bibitem[\protect\citeauthoryear{Cabre \& Gaztanaga}{Cabre \&
  Gaztanaga}{2009}]{Cabre:2008sz}
Cabre A.,  Gaztanaga E.,  2009, \mn@doi [Mon. Not. Roy. Astron. Soc.]
  {10.1111/j.1365-2966.2008.14281.x}, 393, 1183

\bibitem[\protect\citeauthoryear{{Cahn}, {Bailey}, {Dawson}, {Forero Romero},
  {Schlegel}, {White}  \& {DESI}}{{Cahn} et~al.}{2015}]{2015AAS...22533610C}
{Cahn} R.~N.,  {Bailey} S.~J.,  {Dawson} K.~S.,  {Forero Romero} J.,
  {Schlegel} D.~J.,  {White} M.,   {DESI} 2015, in American Astronomical
  Society Meeting Abstracts. p. 336.10

\bibitem[\protect\citeauthoryear{Casaponsa, Heavens, Kitching, Miller, Barreiro
   \& Martinez-Gonzalez}{Casaponsa et~al.}{2013}]{Casaponsa:2012tq}
Casaponsa B.,  Heavens A.~F.,  Kitching T.~D.,  Miller L.,  Barreiro R.~B.,
  Martinez-Gonzalez E.,  2013, \mn@doi [Mon. Not. Roy. Astron. Soc.]
  {10.1093/mnras/stt088}, 430, 2844

\bibitem[\protect\citeauthoryear{Challinor \& Lewis}{Challinor \&
  Lewis}{2011}]{Challinor:2011bk}
Challinor A.,  Lewis A.,  2011, \mn@doi [Phys. Rev.]
  {10.1103/PhysRevD.84.043516}, D84, 043516

\bibitem[\protect\citeauthoryear{Chevallier \& Polarski}{Chevallier \&
  Polarski}{2001}]{Chevallier:2000qy}
Chevallier M.,  Polarski D.,  2001, \mn@doi [Int. J. Mod. Phys.]
  {10.1142/S0218271801000822}, D10, 213

\bibitem[\protect\citeauthoryear{Chuang et~al.}{Chuang
  et~al.}{2013}]{Chuang:2013wga}
Chuang C.-H.,  et~al., 2013

\bibitem[\protect\citeauthoryear{Cresswell \& Percival}{Cresswell \&
  Percival}{2009}]{Cresswell:2008aa}
Cresswell J.~G.,  Percival W.~J.,  2009, \mn@doi [Mon. Not. Roy. Astron. Soc.]
  {10.1111/j.1365-2966.2008.14082.x}, 392, 682

\bibitem[\protect\citeauthoryear{Djorgovski \& Davis}{Djorgovski \&
  Davis}{1987}]{Djorgovski:1987vx}
Djorgovski S.,  Davis M.,  1987, \mn@doi [Astrophys. J.] {10.1086/164948}, 313,
  59

\bibitem[\protect\citeauthoryear{Dressler, Lynden-Bell, Burstein, Davies,
  Faber, Terlevich  \& Wegner}{Dressler et~al.}{1987}]{Dressler:1987ny}
Dressler A.,  Lynden-Bell D.,  Burstein D.,  Davies R.~L.,  Faber S.~M.,
  Terlevich R.,   Wegner G.,  1987, \mn@doi [Astrophys. J.] {10.1086/164947},
  313, 42

\bibitem[\protect\citeauthoryear{Ferreira \& Skordis}{Ferreira \&
  Skordis}{2010}]{Ferreira:2010sz}
Ferreira P.~G.,  Skordis C.,  2010, \mn@doi [Phys. Rev.]
  {10.1103/PhysRevD.81.104020}, D81, 104020

\bibitem[\protect\citeauthoryear{{Fry}}{{Fry}}{1996}]{1996ApJ...461L..65F}
{Fry} J.~N.,  1996, \mn@doi [\apjl] {10.1086/310006}, \href
  {http://adsabs.harvard.edu/abs/1996ApJ...461L..65F} {461, L65}

\bibitem[\protect\citeauthoryear{{Futamase} \& {Sasaki}}{{Futamase} \&
  {Sasaki}}{1989}]{1989PhRvD..40.2502F}
{Futamase} T.,  {Sasaki} M.,  1989, \mn@doi [\prd] {10.1103/PhysRevD.40.2502},
  \href {http://adsabs.harvard.edu/abs/1989PhRvD..40.2502F} {40, 2502}

\bibitem[\protect\citeauthoryear{Gaztanaga, Bonvin  \& Hui}{Gaztanaga
  et~al.}{2015}]{Gaztanaga:2015jrs}
Gaztanaga E.,  Bonvin C.,   Hui L.,  2015

\bibitem[\protect\citeauthoryear{Grieb et~al.}{Grieb
  et~al.}{2016}]{Grieb:2016uuo}
Grieb J.~N.,  et~al., 2016, Submitted to: Mon. Not. Roy. Astron. Soc.

\bibitem[\protect\citeauthoryear{Guzzo et~al.}{Guzzo
  et~al.}{2008}]{Guzzo:2008ac}
Guzzo L.,  et~al., 2008, \mn@doi [Nature] {10.1038/nature06555}, 451, 541

\bibitem[\protect\citeauthoryear{Hall \& Bonvin}{Hall \&
  Bonvin}{2016}]{Hall:2016bmm}
Hall A.,  Bonvin C.,  2016

\bibitem[\protect\citeauthoryear{{Hamilton}}{{Hamilton}}{1992}]{1992ApJ...385L...5H}
{Hamilton} A.~J.~S.,  1992, \mn@doi [\apjl] {10.1086/186264}, \href
  {http://adsabs.harvard.edu/abs/1992ApJ...385L...5H} {385, L5}

\bibitem[\protect\citeauthoryear{Harrison, Camera, Zuntz  \& Brown}{Harrison
  et~al.}{2016}]{Harrison:2016stv}
Harrison I.,  Camera S.,  Zuntz J.,   Brown M.~L.,  2016, ]
  {10.1093/mnras/stw2082}

\bibitem[\protect\citeauthoryear{Hawkins et~al.}{Hawkins
  et~al.}{2003}]{Hawkins:2002sg}
Hawkins E.,  et~al., 2003, \mn@doi [Mon. Not. Roy. Astron. Soc.]
  {10.1046/j.1365-2966.2003.07063.x}, 346, 78

\bibitem[\protect\citeauthoryear{Heavens, Alsing  \& Jaffe}{Heavens
  et~al.}{2013}]{Heavens:2013gol}
Heavens A.,  Alsing J.,   Jaffe A.,  2013, \mn@doi [Mon. Not. Roy. Astron.
  Soc.] {10.1093/mnrasl/slt045}, 433, 6

\bibitem[\protect\citeauthoryear{Hui \& Greene}{Hui \&
  Greene}{2006}]{Hui:2005nm}
Hui L.,  Greene P.~B.,  2006, \mn@doi [Phys. Rev.]
  {10.1103/PhysRevD.73.123526}, D73, 123526

\bibitem[\protect\citeauthoryear{Jeong, Schmidt  \& Hirata}{Jeong
  et~al.}{2012}]{Jeong:2011as}
Jeong D.,  Schmidt F.,   Hirata C.~M.,  2012, \mn@doi [Phys. Rev.]
  {10.1103/PhysRevD.85.023504}, D85, 023504

\bibitem[\protect\citeauthoryear{Kaiser}{Kaiser}{1987}]{Kaiser:1987qv}
Kaiser N.,  1987, Mon. Not. Roy. Astron. Soc., 227, 1

\bibitem[\protect\citeauthoryear{Kaiser \& Hudson}{Kaiser \&
  Hudson}{2015}]{Kaiser:2014jca}
Kaiser N.,  Hudson M.~J.,  2015, \mn@doi [Mon. Not. Roy. Astron. Soc.]
  {10.1093/mnras/stv693}, 450, 883

\bibitem[\protect\citeauthoryear{{Kasai}, {Futamase}  \& {Takahara}}{{Kasai}
  et~al.}{1990}]{1990PhLA..147...97K}
{Kasai} M.,  {Futamase} T.,   {Takahara} F.,  1990, \mn@doi [Physics Letters A]
  {10.1016/0375-9601(90)90875-O}, \href
  {http://adsabs.harvard.edu/abs/1990PhLA..147...97K} {147, 97}

\bibitem[\protect\citeauthoryear{Levi et~al.}{Levi et~al.}{2013}]{Levi:2013gra}
Levi M.,  et~al., 2013

\bibitem[\protect\citeauthoryear{{Lilje} \& {Efstathiou}}{{Lilje} \&
  {Efstathiou}}{1989}]{1989MNRAS.236..851L}
{Lilje} P.~B.,  {Efstathiou} G.,  1989, \mnras, \href
  {http://adsabs.harvard.edu/abs/1989MNRAS.236..851L} {236, 851}

\bibitem[\protect\citeauthoryear{Linder}{Linder}{2003}]{Linder:2002et}
Linder E.~V.,  2003, \mn@doi [Phys. Rev. Lett.]
  {10.1103/PhysRevLett.90.091301}, 90, 091301

\bibitem[\protect\citeauthoryear{Linder}{Linder}{2005}]{Linder:2005in}
Linder E.~V.,  2005, \mn@doi [Phys. Rev.] {10.1103/PhysRevD.72.043529}, D72,
  043529

\bibitem[\protect\citeauthoryear{Montanari \& Durrer}{Montanari \&
  Durrer}{2012}]{Montanari:2012me}
Montanari F.,  Durrer R.,  2012, \mn@doi [Phys. Rev.]
  {10.1103/PhysRevD.86.063503}, D86, 063503

\bibitem[\protect\citeauthoryear{{Nusser} \& {Davis}}{{Nusser} \&
  {Davis}}{1994}]{1994ApJ...421L...1N}
{Nusser} A.,  {Davis} M.,  1994, \mn@doi [\apjl] {10.1086/187172}, \href
  {http://adsabs.harvard.edu/abs/1994ApJ...421L...1N} {421, L1}

\bibitem[\protect\citeauthoryear{Papai \& Szapudi}{Papai \&
  Szapudi}{2008}]{Papai:2008bd}
Papai P.,  Szapudi I.,  2008, \mn@doi [Mon. Not. Roy. Astron. Soc.]
  {10.1111/j.1365-2966.2008.13572.x}, 389, 292

\bibitem[\protect\citeauthoryear{Percival et~al.}{Percival
  et~al.}{2007}]{Percival:2006gt}
Percival W.~J.,  et~al., 2007, \mn@doi [Astrophys. J.] {10.1086/510615}, 657,
  645

\bibitem[\protect\citeauthoryear{Pyne \& Birkinshaw}{Pyne \&
  Birkinshaw}{2004}]{Pyne:2003bn}
Pyne T.,  Birkinshaw M.,  2004, \mn@doi [Mon. Not. Roy. Astron. Soc.]
  {10.1111/j.1365-2966.2004.07362.x}, 348, 581

\bibitem[\protect\citeauthoryear{Samushia et~al.}{Samushia
  et~al.}{2014}]{Samushia:2013yga}
Samushia L.,  et~al., 2014, \mn@doi [Mon. Not. Roy. Astron. Soc.]
  {10.1093/mnras/stu197}, 439, 3504

\bibitem[\protect\citeauthoryear{{Sasaki}}{{Sasaki}}{1987}]{1987MNRAS.228..653S}
{Sasaki} M.,  1987, \mn@doi [\mnras] {10.1093/mnras/228.3.653}, \href
  {http://adsabs.harvard.edu/abs/1987MNRAS.228..653S} {228, 653}

\bibitem[\protect\citeauthoryear{Satpathy et~al.}{Satpathy
  et~al.}{2016}]{Satpathy:2016tct}
Satpathy S.,  et~al., 2016, Submitted to: Mon. Not. Roy. Astron. Soc.

\bibitem[\protect\citeauthoryear{{Schmidt}, {Leauthaud}, {Massey}, {Rhodes},
  {George}, {Koekemoer}, {Finoguenov}  \& {Tanaka}}{{Schmidt}
  et~al.}{2012}]{2012ApJ...744L..22S}
{Schmidt} F.,  {Leauthaud} A.,  {Massey} R.,  {Rhodes} J.,  {George} M.~R.,
  {Koekemoer} A.~M.,  {Finoguenov} A.,   {Tanaka} M.,  2012, \mn@doi [\apjl]
  {10.1088/2041-8205/744/2/L22}, \href
  {http://adsabs.harvard.edu/abs/2012ApJ...744L..22S} {744, L22}

\bibitem[\protect\citeauthoryear{Song, Sabiu, Kayo  \& Nichol}{Song
  et~al.}{2011}]{Song:2010kq}
Song Y.-S.,  Sabiu C.~G.,  Kayo I.,   Nichol R.~C.,  2011, \mn@doi [JCAP]
  {10.1088/1475-7516/2011/05/020}, 1105, 020

\bibitem[\protect\citeauthoryear{Springob et~al.,}{Springob
  et~al.}{2014}]{Springob:2014qja}
Springob C.~M.,  et~al., 2014, \mn@doi [Mon. Not. Roy. Astron. Soc.]
  {10.1093/mnras/stu1743}, 445, 2677

\bibitem[\protect\citeauthoryear{Szalay, Matsubara  \& Landy}{Szalay
  et~al.}{1998}]{Szalay:1997cc}
Szalay A.~S.,  Matsubara T.,   Landy S.~D.,  1998, \mn@doi [Astrophys. J.]
  {10.1086/311293}, 498, L1

\bibitem[\protect\citeauthoryear{Szapudi}{Szapudi}{2004}]{Szapudi:2004gh}
Szapudi I.,  2004, \mn@doi [Astrophys. J.] {10.1086/423168}, 614, 51

\bibitem[\protect\citeauthoryear{Tegmark \& Peebles}{Tegmark \&
  Peebles}{1998}]{Tegmark:1998wm}
Tegmark M.,  Peebles P. J.~E.,  1998, \mn@doi [Astrophys. J.] {10.1086/311426},
  500, L79

\bibitem[\protect\citeauthoryear{Tonry, Blakeslee, Ajhar  \& Dressler}{Tonry
  et~al.}{2000}]{Tonry:1999vt}
Tonry J.~L.,  Blakeslee J.~P.,  Ajhar E.~A.,   Dressler A.,  2000, \mn@doi
  [Astrophys. J.] {10.1086/308409}, 530, 625

\bibitem[\protect\citeauthoryear{{Tully} \& {Fisher}}{{Tully} \&
  {Fisher}}{1977}]{1977A&A....54..661T}
{Tully} R.~B.,  {Fisher} J.~R.,  1977, \aap, \href
  {http://adsabs.harvard.edu/abs/1977A%26A....54..661T} {54, 661}

\bibitem[\protect\citeauthoryear{Tully et~al.}{Tully
  et~al.}{2013}]{Tully:2013wqa}
Tully R.~B.,  et~al., 2013, \mn@doi [Astron. J.] {10.1088/0004-6256/146/4/86},
  146, 86

\bibitem[\protect\citeauthoryear{Turnbull, Hudson, Feldman, Hicken, Kirshner
  \& Watkins}{Turnbull et~al.}{2012}]{Turnbull:2011ty}
Turnbull S.~J.,  Hudson M.~J.,  Feldman H.~A.,  Hicken M.,  Kirshner R.~P.,
  Watkins R.,  2012, \mn@doi [Mon. Not. Roy. Astron. Soc.]
  {10.1111/j.1365-2966.2011.20050.x}, 420, 447

\bibitem[\protect\citeauthoryear{Wang \& Steinhardt}{Wang \&
  Steinhardt}{1998}]{Wang:1998gt}
Wang L.-M.,  Steinhardt P.~J.,  1998, \mn@doi [Astrophys. J.] {10.1086/306436},
  508, 483

\bibitem[\protect\citeauthoryear{Yoo}{Yoo}{2010}]{Yoo:2010ni}
Yoo J.,  2010, \mn@doi [Phys. Rev.] {10.1103/PhysRevD.82.083508}, D82, 083508

\bibitem[\protect\citeauthoryear{Yoo, Fitzpatrick  \& Zaldarriaga}{Yoo
  et~al.}{2009}]{Yoo:2009au}
Yoo J.,  Fitzpatrick A.~L.,   Zaldarriaga M.,  2009, \mn@doi [Phys. Rev.]
  {10.1103/PhysRevD.80.083514}, D80, 083514

\bibitem[\protect\citeauthoryear{Zehavi et~al.,}{Zehavi
  et~al.}{2005}]{Zehavi:2004zn}
Zehavi I.,  et~al., 2005, \mn@doi [Astrophys. J.] {10.1086/427495}, 621, 22

\makeatother
\end{thebibliography}

\bsp	
\label{lastpage}
\end{document}